\documentclass[twocolumn,showpacs,aps,prd]{revtex4}

\usepackage{graphicx}
\usepackage{dcolumn}
\usepackage{amsmath}
\usepackage{epsfig}
\usepackage{diagbox}
\usepackage{url}

\input{pubboard/babarsym.tex}

\def\ie                 {{\it i.e.}}

\def\mkspip 		{\mbox{$m_{\KS\pip}$}}
\def\mkspim 		{\mbox{$m_{\KS\pim}$}}
\def\mkspiz 		{\mbox{$m_{\KS\piz}$}}
\def\mpippiz 		{\mbox{$m_{\pip\piz}$}}
\def\mpimpiz 		{\mbox{$m_{\pim\piz}$}}

\def\mkspipSq 		{\mbox{$m^{2}_{\KS\pip}$}}
\def\mkspizSq 		{\mbox{$m^{2}_{\KS\piz}$}}
\def\mpippizSq 		{\mbox{$m^{2}_{\pip\piz}$}}

\def\mkspimSq 		{\mbox{$m^{2}_{\KS\pim}$}}
\def\mpimpizSq 		{\mbox{$m^{2}_{\pim\piz}$}}

\def\Acp		{\mbox{$A_{\CP}$}}

\def\D       {\ensuremath{D}\xspace}

\newcommand{\laura}               {\mbox{\tt Laura++}\xspace}
\newcommand{\rhop}               {\mbox{$\rho^+$}}

\newcommand{\KstarIp}             {\mbox{$\Kstar(892)^{+}$}}
\newcommand{\KstarIm}             {\mbox{$\Kstar(892)^{-}$}}
\newcommand{\KstarIpm}             {\mbox{$\Kstar(892)^{\pm}$}}
\newcommand{\BptoKstarIppiz}      {\mbox{$\Bp \to \KstarIp\piz$}}
\newcommand{\BmtoKstarImpiz}      {\mbox{$\Bm \to \KstarIm\piz$}}
\newcommand{\KstarI}             {\mbox{$\Kstar(892)^{0}$}}
\newcommand{\BptoKstarIpip}      {\mbox{$\Bp \to \KstarI\pip$}}
\newcommand{\KstarIIp}            {\mbox{$\Kstar_{0}(1430)^{+}$}}
\newcommand{\BptoKstarIIppiz}      {\mbox{$\Bp \to \KstarIIp\piz$}}
\newcommand{\KstarII}            {\mbox{$\Kstar_{0}(1430)^{0}$}}
\newcommand{\BptoKstarIIpip}      {\mbox{$\Bp \to \KstarII\pip$}}
\newcommand{\rhoIp}               {\mbox{$\rho(770)^{+}$}}
\newcommand{\rhoIm}               {\mbox{$\rho(770)^{-}$}}
\newcommand{\rhoIpm}               {\mbox{$\rho(770)^{\pm}$}}
\newcommand{\KpiSwave}           {\mbox{$(K\pi)^{*0}_0$}}
\newcommand{\KpiSwavep}        	{\mbox{$(K\pi)^{*+}_0$}}
\newcommand{\KpiSwavem}        	{\mbox{$(K\pi)^{*-}_0$}}
\newcommand{\KpiSwavepm}        	{\mbox{$(K\pi)^{*\pm}_0$}}
\newcommand{\BptoKpiSwavepip}      {\mbox{$\Bp \to \KpiSwave\pip$}}
\newcommand{\BptoKpiSwaveppiz}     {\mbox{$\Bp \to \KpiSwavep\piz$}}

\newcommand{\BptorhopKs}           {\mbox{$\Bp \to \rhoIp\KS$}}
\newcommand{\BptorhopKz}           {\mbox{$\Bp \to \rhoIp\Kz$}}
\newcommand{\BptoKspippiz}        {\mbox{$\Bp \to \KS \pip \piz$}}

\newcommand{\BptoKzpippiz}        {\mbox{$\Bp \to \Kz \pip \piz$}}
\newcommand{\Kspippiz}        {\mbox{$\KS \pip \piz$}}
\newcommand{\BptoKppimpip}		{\mbox{$\Bp \to \Kp\pim\pip$}}
\newcommand{\BptoKppizpiz}       {\mbox{$\Bp \to \Kp\piz\piz$}}

\newcommand{\KstarIII}           {\mbox{$\Kstar_{2}(1430)^{0}$}}
\newcommand{\KstarIIIp}           {\mbox{$\Kstar_{2}(1430)^{+}$}}
\newcommand{\KstarIV}            {\mbox{$\Kstar(1680)^{0}$}}
\newcommand{\KstarIVp}            {\mbox{$\Kstar(1680)^{+}$}}
\newcommand{\rhoIIp}              {\mbox{$\rho(1450)^{+}$}}

\def\ncand{\ensuremath{31\,876}}
\def\nsig{\ensuremath{1014 \pm 60}}

\def\kpipiBFal{\ensuremath{45.9 \pm 2.6 \pm 3.0^{+8.6}_{-0.0}}}
\def\kpipiAcp{\ensuremath{0.07 \pm 0.05 \pm 0.03^{+0.02}_{-0.03}}}

\def\kstarIpipBF{\ensuremath{14.6 \pm 2.4 \pm 1.4^{+0.3}_{-0.4}}}
\def\kstarIpipAcp{\ensuremath{-0.12 \pm 0.21 \pm 0.08^{+0.0}_{-0.11}}}

\def\kstarIppizBF{\ensuremath{9.2 \pm 1.3 \pm 0.6^{+0.3}_{-0.5}}}
\def\kstarIppizAcp{\ensuremath{-0.52 \pm 0.14 \pm 0.04^{+0.04}_{-0.02}}}

\def\kstarIIpipBF{\ensuremath{50.0 \pm 4.8 \pm 6.1^{+2.7}_{-2.6}}}
\def\kstarIIpipAcp{\ensuremath{0.14 \pm 0.10 \pm 0.04^{+0.13}_{-0.05}}}

\def\kstarIIppizBF{\ensuremath{17.2 \pm 2.4 \pm 1.5^{+0.0}_{-1.8}}}
\def\kstarIIppizAcp{\ensuremath{0.26 \pm 0.12 \pm 0.08^{+0.12}_{-0.0}}}

\def\rhoIpKzBF{\ensuremath{9.4 \pm 1.6 \pm 1.1^{+0.0}_{-2.6}}}
\def\rhoIpKzAcp{\ensuremath{0.21 \pm 0.19 \pm 0.07^{+0.23}_{-0.19}}}

\def\kstarIpipBFComb{\ensuremath{11.6 \pm 0.5 \pm 1.1}}
\def\kstarIpipAcpComb{\ensuremath{0.025 \pm 0.050 \pm 0.016}}

\def\kstarIppizBFComb{\ensuremath{8.8 \pm 1.0 \pm 0.6}}
\def\kstarIppizAcpComb{\ensuremath{-0.39 \pm 0.12 \pm 0.03}}

\def\kstarIIpipBFComb{\ensuremath{43.4 \pm 1.2 \pm 5.8}}
\def\kstarIIpipAcpComb{\ensuremath{0.040 \pm 0.033 \pm 0.033}}

\def\kstarIppizPhNonCP{\ensuremath{-95 \pm 43^{+48\,+8}_{-36\,-70}}}
\def\kstarIIpipPhNonCP{\ensuremath{174 \pm 11 \pm 11^{+0}_{-6}}}
\def\kstarIIppizPhNonCP{\ensuremath{-89 \pm 43^{+53\,+5}_{-40\,-17}}}
\def\rhoIpKzPhNonCP{\ensuremath{-122 \pm 43^{+55\,+16}_{-47\,-66}}}

\def\kstarIppizA{\ensuremath{1.46 \pm 0.22 \pm 0.05^{+0.05}_{-0.03}}}
\def\kstarIIpipA{\ensuremath{1.74 \pm 0.21 \pm 0.11^{+0.07}_{-0.12}}}
\def\kstarIIppizA{\ensuremath{1.44 \pm 0.22 \pm 0.13^{+0.00}_{-0.10}}}
\def\rhoIpKzA{\ensuremath{1.24 \pm 0.01 \pm 0.09^{+0.00}_{-0.21}}}

\def\kstarIppizAbar{\ensuremath{0.82 \pm 0.18 \pm 0.05^{+0.06}_{-0.04}}}
\def\kstarIIpipAbar{\ensuremath{2.00 \pm 0.27 \pm 0.13^{+0.14}_{-0.02}}}
\def\kstarIIppizAbar{\ensuremath{1.88 \pm 0.25 \pm 0.14^{+0.22}_{-0.06}}}
\def\rhoIpKzAbar{\ensuremath{1.54 \pm 0.01 \pm 0.09^{+0.23}_{-0.09}}}

\def\kstarIppizPh{\ensuremath{-10 \pm 112}}
\def\kstarIIpipPh{\ensuremath{165 \pm 19 \pm 9^{+4}_{-3}}}
\def\kstarIIppizPh{\ensuremath{4 \pm 111}}
\def\rhopKzPh{\ensuremath{-50 \pm 168}}

\def\kstarIppizPhBar{\ensuremath{-98 \pm 97}}
\def\kstarIIpipPhBar{\ensuremath{190 \pm 21 \pm 11^{+1}_{-3}}}
\def\kstarIIppizPhBar{\ensuremath{-109 \pm 92}}
\def\rhopKzPhBar{\ensuremath{-120 \pm 71}}

\def\deltaKstarppizPh{\ensuremath{-14 \pm 18 \pm 9^{+4}_{-3}}}
\def\deltaKstarppizPhBar{\ensuremath{11 \pm 19 \pm 10^{+17}_{-9}}}

\newcommand{\onreslumi}  {\mbox{429\invfb}}
\newcommand{\offreslumi} {\mbox{45\invfb}}
\newcommand{\bbpairs}    {\mbox{$470.9\pm2.8$~million}}

\newcommand{\nbb}        {\mbox{$N_{\BB}$}}
\newcommand{\NN}         {\mbox{${\rm BDT}_{\rm out}$}}

\newcommand{\BABARPubYear}    {14}
\newcommand{\BABARPubNumber}  {011}
\newcommand{\SLACPubNumber} {16186}

\begin{document}

\preprint{\babar-PUB-\BABARPubYear/\BABARPubNumber} 
\preprint{SLAC-PUB-\SLACPubNumber} 

\begin{flushleft}
  \babar-PUB-\BABARPubYear/\BABARPubNumber\\
  SLAC-PUB-\SLACPubNumber\\[10mm]
\end{flushleft}

\title{
  {\large 
    \bf Evidence for \CP\ violation in \BptoKstarIppiz\ from a Dalitz plot
    analysis of \BptoKspippiz\ decays}
}

%
\author{J.~P.~Lees}
\author{V.~Poireau}
\author{V.~Tisserand}
\affiliation{Laboratoire d'Annecy-le-Vieux de Physique des Particules (LAPP), Universit\'e de Savoie, CNRS/IN2P3,  F-74941 Annecy-Le-Vieux, France}
\author{E.~Grauges}
\affiliation{Universitat de Barcelona, Facultat de Fisica, Departament ECM, E-08028 Barcelona, Spain }
\author{A.~Palano$^{ab}$ }
\affiliation{INFN Sezione di Bari$^{a}$; Dipartimento di Fisica, Universit\`a di Bari$^{b}$, I-70126 Bari, Italy }
\author{G.~Eigen}
\author{B.~Stugu}
\affiliation{University of Bergen, Institute of Physics, N-5007 Bergen, Norway }
\author{D.~N.~Brown}
\author{L.~T.~Kerth}
\author{Yu.~G.~Kolomensky}
\author{M.~J.~Lee}
\author{G.~Lynch}
\affiliation{Lawrence Berkeley National Laboratory and University of California, Berkeley, California 94720, USA }
% ...
\collaboration{The \babar\ Collaboration}
\noaffiliation

\begin{abstract}
	We report a Dalitz plot analysis of charmless hadronic decays of
	charged \B\ mesons to the final state \Kspippiz\ using the full
	\babar\ dataset of \bbpairs\ \BB\ events collected at the \FourS\
	resonance.  We measure the overall branching fraction and \CP\ asymmetry to
	be ${\cal B}\left(\BptoKzpippiz\right) =
	\left(\kpipiBFal\right)\times10^{-6}$ and
	$\Acp\left(\BptoKzpippiz\right) = \kpipiAcp$, where the
	uncertainties are statistical, systematic, and due to the signal
	model, respectively. This is the first measurement of the branching
	fraction for \BptoKzpippiz.  We find first evidence of a \CP\
	asymmetry in \BptoKstarIppiz\ decays:
	$\Acp\left(\BptoKstarIppiz\right)=\kstarIppizAcp$.  The
	significance of this asymmetry, including systematic and model
	uncertainties, is $3.4$ standard deviations.  We also
	measure the branching fractions and \CP\ asymmetries for three
	other intermediate decay modes.
\end{abstract}

\pacs{13.25.Hw, 12.15.Hh, 11.30.Er}

\maketitle


\section{Introduction}

The Cabibbo-Kobayashi-Maskawa (CKM) mechanism~\cite{Cabibbo:1963yz,Kobayashi:1973fv} for quark mixing
describes all weak charged current transitions between quarks in terms of a unitarity matrix
with four parameters: three
rotation angles and an irreducible phase. The unitarity of the CKM matrix
is usually expressed as triangle relationships among its elements.
The interference between tree-level and loop (``penguin'') amplitudes can give rise to
direct \CP\ violation, which is sensitive to the angles of the Unitarity Triangle,
denoted $\alpha$, $\beta$, and $\gamma$. Measurements of the parameters of
the CKM matrix provide an important test of the Standard Model (SM) since
any deviation from unitarity or discrepancies between measurements of the
same parameter in different decay processes would imply a possible
signature of new physics.  Tree amplitudes in $\B\to\Kstar\pi$ decays are
sensitive to $\gamma$, which can be extracted from interferences between the
intermediate states that populate the $K\pi\pi$ Dalitz plane. However,
these amplitudes are Cabibbo-suppressed relative to contributions carrying
a different phase and involving radiation of either a gluon (QCD penguin)
or photon (electroweak penguin or EWP) from a loop.

QCD penguin contributions can be eliminated by
constructing a linear combination of the weak decay amplitudes for
$\Bp\to\Kstar\pi$ to form a pure isospin
$I=\frac{3}{2}$ state~\cite{Ciuchini:2006kv}: 
\begin{equation}
	A_{\frac{3}{2}}	= A\left(K^{*0}\pi^{+}\right)+\sqrt{2}A\left(K^{*+}\pi^{0}\right).
	\label{eq:amp-rel}
\end{equation}
Since all transitions from $I=\frac{1}{2}$ to $I=\frac{3}{2}$ states occur
via only $\Delta I=1$ operators, $A_{\frac{3}{2}}$ is free from
QCD contributions. The weak phase of $A_{\frac{3}{2}}$ is often denoted as
\begin{equation}
\Phi_{\frac{3}{2}}=-\frac{1}{2}{\rm Arg}\left(\bar{A}_{\frac{3}{2}}/A_{\frac{3}{2}}\right),
\label{eq:phase-rel}
\end{equation}
where $\bar{A}_{\frac{3}{2}}$ is the \CP\ conjugate of the amplitude in
Eq.~(\ref{eq:amp-rel}). The phase $\Phi_{\frac{3}{2}}$ in
Eq.~(\ref{eq:phase-rel}) is the CKM angle $\gamma$ in the absence of EWP
contributions~\cite{Gronau:2006qn}. 

Measurements of the rates and \CP\ asymmetries in $\B\to K\pi$ have
generated considerable interest because of possible hints of new-physics
contributions~\cite{Aubert:2007hh,:2008zza}. Of particular interest is the
difference, $\Delta\Acp$, between the \CP\ asymmetry in $\Bp\to\Kp\piz$ and
the \CP\ asymmetry in $\Bz\to\Kp\pim$, which in the SM is expected to be
consistent with zero within the theoretical uncertainties assuming U-spin
symmetry and in the absence of color-suppressed tree and electroweak
amplitudes~\cite{Gronau:2005kz,Buras:2003dj}. Using the
average values of \Acp\ of $\Kp\piz$ and $\Kp\pim$
decays~\cite{Amhis:2012bh}, $\Delta\Acp\left(K\pi\right)$ is 
\begin{eqnarray}
	\Delta\Acp\left(K\pi\right)&=&\Acp\left(\Kp\piz\right)-\Acp\left(\Kp\pim\right)
	\nonumber \\
				   &=&0.122\pm0.022,
\label{eq:deltaAcp-Kpi}
\end{eqnarray}
which differs from zero by $5.5$ standard deviations.
Unfortunately, hadronic uncertainties prevent a clear interpretation of
these results in terms of the new-physics
implications~\cite{Gronau:2008gu,Ciuchini:2006kv}.  Additional 
information can be obtained through studies of the related
vector-pseudoscalar decays $B\to \Kstar\pi$ and $B\to
K\rho$~\cite{Chang:2008tf,Chiang:2009hd,Gronau:2010dd},  for which the ratios of
tree-to-penguin amplitudes are expected to be two to three times larger
than for $B\to K\pi$ decays.  Hence, $B\to \Kstar\pi$ and $B\to K\rho$
decays could have considerably larger \CP\ asymmetries.

In this article, we present the results from an amplitude analysis of
\BptoKspippiz\ decays. The inclusion of charge conjugate processes is
implied throughout this article, except when referring to \CP\ asymmetries.
This is the first Dalitz plot analysis of this decay by \babar; the only
previous \babar\ analysis of this decay was restricted to measuring the
branching fraction and \CP\ asymmetry of
$\Bp\to\Kz\rhop$~\cite{Aubert:2007mb}.  An upper limit on the branching
fraction for \BptoKzpippiz\ was set by the CLEO Collaboration: ${\cal
B}(\BptoKzpippiz)<66\times10^{-6}$~\cite{Eckhart:2002qr}. 

Two contributions to the \Kspippiz\ final state arise from the resonant
decays \BptoKstarIpip\ and \BptoKstarIppiz. Although both the rate and \CP\
asymmetries for \BptoKstarIpip\ have been well measured, with
$\Kstarz\to\Kp\pim$, by both the \babar~\cite{Aubert:2008bj} and
Belle~\cite{Garmash:2005rv} Collaborations, the measurements of the rate
and \CP\ asymmetry for \BptoKstarIppiz~\cite{BABAR:2011aaa} have
significant statistical uncertainties
and could benefit from the additional information provided by a full
amplitude analysis.  In \tabref{status} we review the existing 
measurements of the rates and \CP\ asymmetries in the $B\to \Kstar(892)\pi$ system. 
\begin{table}[htb]
\center
\caption{
	Average values of the branching fractions ${\cal B}$ and \CP\
	asymmetries \Acp\ for $B\to \Kstar(892)\pi$ decays as determined by
	the Heavy Flavor Averaging Group~\cite{Amhis:2012bh}.  
	}
\label{tab:status}
\begin{tabular}{c@{\hspace{5mm}}c@{\hspace{5mm}}c@{\hspace{5mm}}c}
\hline
Mode & ${\cal B}(10^{-6})$ & \Acp & References \\
\hline
\phantom{\huge I}$\Kstarp\pim$ & $8.5\pm 0.7$ & $-0.23 \pm 0.06$ &
\cite{BABAR:2011ae,Aubert:2009me,Dalseno:2008wwa,Garmash:2006fh} \\
$\Kstarp\piz$ & $8.2\pm 1.8$ & $-0.06\pm0.24$ & \cite{BABAR:2011aaa} \\
$\Kstarz\pip$ & $9.9\,^{+0.8}_{-0.9}$ & $-0.038\pm0.042$ &
\cite{Aubert:2008bj,Garmash:2005rv} \\
$\Kstarz\piz$ & $2.5 \pm 0.6$ & $-0.15 \pm 0.13$ & \cite{Chang:2004um,BABAR:2011ae} \\
\hline
\end{tabular}
\end{table}

This article is organised as follows. The isobar model used to parameterize
the complex amplitudes describing the intermediate resonances contributing
to the $\KS\pip\piz$ final state is presented in
\secref{dp-model}. A brief description of the \babar\ detector and the
dataset is given in \secref{detector}. The event reconstruction and
selection are discussed in detail in \secref{event-reco}, the background
study in \secref{bkg-study}, and a description of the extended maximum
likelihood fit in \secref{max-like}. The results are given in
\secref{results}, and a study of the systematic uncertainties is presented in
\secref{systematics}. In \secref{conclusion}, we provide a summary and
conclusion, discussing the results and combining the branching fractions
and \CP\ asymmetries for the decays \BptoKstarIpip, \BptoKstarIIpip, and
\BptoKstarIppiz\ with previous \babar\ results obtained from the final
states \BptoKppimpip\ and \BptoKppizpiz.

\section{Amplitude analysis formalism}
\label{sec:dp-model}

A number of intermediate states contribute to the decay \BptoKspippiz.
Their individual contributions are measured by performing a maximum
likelihood fit to the distribution of events in the Dalitz plot formed from
the two variables, $m^{2}_{\KS\pip}$ and $m^{2}_{\pip\piz}$. We use the
\laura~\cite{laurapp} software to perform this fit.

The total signal amplitudes for the \Bp\ and the \Bm\ decays are given in
the isobar formalism by~\cite{Fleming:1964zz,Herndon:1973yn}
\small
\begin{eqnarray}
	A\left(\mkspipSq,\mpippizSq\right)=\sum_{j}c_{j}F_{j}\left(\mkspipSq,\mpippizSq\right)\label{isobar-formalism-A}, \\
  \bar{A}\left(\mkspimSq,\mpimpizSq\right)=\sum_{j}\bar{c}_{j}\bar{F}_{j}\left(\mkspimSq,\mpimpizSq\right)\label{isobar-formalism-Abar}, 
\end{eqnarray}
\normalsize
where $c_{j}$ is the complex coefficient for a given resonant decay mode
$j$ contributing to the Dalitz plot. This complex coefficient contains the
weak-interaction phase dependence that is measured relative to one of the contributing
resonant channels. In this article we report results for the relative phases
between each pair of amplitudes.

The function $F_{j}$ describes the dynamics of the decay amplitudes and
is the product of a resonant lineshape ($R_{j}$), two Blatt-Weisskopf
barrier factors~\cite{blatt:weisskopf} ($X_{L}$), and an
angular-dependent term ($T_{j,L}$)~\cite{Beringer:1900zz}:
\begin{equation}
F_{j}=R_{j}\times X_{L}(\left|\vec{p}\right|,\left|\vec{p}_{0}\right|) \times
	X_{L}(\left|\vec{q}\right|,\left|\vec{q}_{0}\right|) \times 
	T_{j,L}\left(\vec{p},\vec{q}\right),
\end{equation}
where $L$ is the orbital angular momentum between the intermediate
resonance and the bachelor particle (the bachelor particle is the daughter of the \B\ decay
that does not arise from the resonance), $\vec{q}$ is the momentum of one of the
daughters of the resonance in the rest frame of the resonance, 
$\vec{p}$ is the momentum of the bachelor particle in the rest frame of the
resonance, and $\vec{p}_{0}$ and $\vec{q}_{0}$ are the values of $\vec{p}$
and $\vec{q}$, respectively, at the nominal mass of the resonance. The
Blatt-Weisskopf barrier factors are given by
\begin{eqnarray}
	X_{L=0}(\left|\vec{u}\right|,\left|\vec{u}_{0}\right|)&=&1,\\
	X_{L=1}(\left|\vec{u}\right|,\left|\vec{u}_{0}\right|)&=&\sqrt{\frac{1+z_{0}}{1+z}},\\
	X_{L=2}(\left|\vec{u}\right|,\left|\vec{u}_{0}\right|)&=&\sqrt{\frac{\left(z_{0}-3\right)^{2}+9z_{0}}{\left(z-3\right)^{2}+9z}},
\end{eqnarray}
where $z=(\left|\vec{u}\right|r_{\rm BW})^{2}$,
$z_{0}=(\left|\vec{u}_{0}\right|r_{\rm BW})^{2}$, $\vec{u}$ is either
$\vec{q}$ or $\vec{p}$, and $r_{\rm BW}=4.0\left(\gevc\right)^{-1}$ is the
meson radius parameter. The uncertainty in $r_{\rm BW}$, used for
systematic variations, is $\pm2~\left(\gevc\right)^{-1}$ for the $K^{*}$
resonances, and ranges from $-1.0$ to $+2.0~\left(\gevc\right)^{-1}$ for
the \rhoIp~\cite{Beringer:1900zz}.  The angular term depends on the spin of
the resonance and is given by~\cite{Zemach:1965zz,Bevan:2014iga}
\begin{eqnarray}
	T_{j,L=0} &=& 1,\\
	T_{j,L=1} &=& -2\vec{p}.\vec{q},\\
	T_{j,L=2} &=&
	\frac{4}{3}\left[3\left(\vec{p}.\vec{q}\right)^{2}-\left(\left|\vec{p}\right|\left|\vec{q}\right|\right)^{2}\right].
\end{eqnarray}

\begin{figure}
	\includegraphics[width=0.5\textwidth]{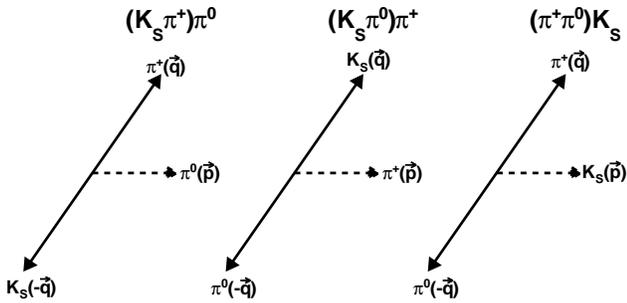}
	\caption{Schematic representation of the definitions of $\vec{q}$
		and $\vec{p}$ used in this analysis for the (left) $\KS\pip$,
		(center) $\KS\piz$, and (right) $\pip\piz$ resonances.
	}
	\label{fig:phase-conv}
\end{figure}

The choice of which resonance daughter is defined to carry the
momentum $\vec{q}$ is a matter of convention. However, its definition is
important when comparing measurements from different experiments.  In \figref{phase-conv}, we
illustrate the momentum definitions used for the $\KS\pip$, $\KS\piz$, and
$\pip\piz$ resonance combinations.

\tabref{signal-model-description} lists the resonances used to model the
signal.  We determine a nominal model from data by studying changes in the
log likelihood values for the best fit when omitting or adding a resonance
to the fit model, as described in \secref{max-like}.

\begin{table}
	\caption{
		Parameters of the Dalitz plot model for \BptoKspippiz\ used
		in the nominal fit. The mass and width of the $\rhoIp$ and
		their uncertainties are taken from the analyses by the
		ALEPH~\cite{Barate:1997hv} and
		CMD2~\cite{Akhmetshin:2001ig} Collaborations. All other parameters are
		taken from Ref.~\cite{Beringer:1900zz}.  The
		resonance shapes are a Gounaris-Sakurai (GS) function, a relativistic
		Breit-Wigner (RBW) function, or based on measurements by the LASS
		Collaboration~\cite{Aston:1987ir}, with $a$ the scattering
		length and $r$ the effective range of the LASS parametrization.
	}
	\label{tab:signal-model-description}
	\begin{tabular}{ll|cc}
		\hline
Resonance & Lineshape 	& \multicolumn{2}{c}{Parameters} \\
	  &		& Resonance mass  & Width 	\\
	  &		& $\left(\mevcc\right)$ 	  & $\left(\mev\right)$ \\
\hline
\phantom{\huge I}\rhoIp			& GS			& $775.5\pm0.6$	& $148.2\pm0.8$	\\
\KstarIp		& RBW		& $891.7\pm0.3$	& $50.8\pm0.9$	\\
\KstarI			& RBW		& $896.1\pm0.2$		& $50.7\pm0.6$	\\
$(K\pi)_{0}^{*0/+}$	& LASS		& $1412\pm50$		& $294\pm80$	\\
			&		& \multicolumn{2}{c}{$m_{{\rm
			cutoff}}=1800\mevcc$~\cite{Aubert:2008bj}} \\
			&		&
			\multicolumn{2}{c}{$a=2.1\pm0.1~(\gevc)^{-1}$~\cite{Aubert:2008bj}} \\
			&		&
			\multicolumn{2}{c}{$r=3.3\pm0.3~(\gevc)^{-1}$~\cite{Aubert:2008bj}} \\
\hline
\end{tabular}
\end{table}

For the \KstarI\ and \KstarIp\ resonances, we use a relativistic
Breit-Wigner (RBW) lineshape~\cite{Beringer:1900zz}:
\begin{equation}
	R_{j}^{\rm RBW}(m)=\frac{1}{m_{0}^{2}-m^{2}-im_{0}\Gamma(m)},
\end{equation}
where $m$ is the two-body invariant mass and $\Gamma(m)$ is the
mass-dependent width. In general, for a resonance decaying to spin-0
particles, $\Gamma(m)$ can be expressed as
\begin{equation}
	\Gamma(m)=\Gamma_{0}\left(\frac{\left|\vec{q}\right|}{\left|\vec{q}_{0}\right|}\right)^{2L+1}\left(\frac{m_{0}}{m}\right)X_{L}(\left|\vec{q}\right|,\left|\vec{q}_{0}\right|)^{2},
	\label{gamma-mass-dep}
\end{equation}
where $m_{0}$ and $\Gamma_{0}$ are the nominal mass and width of the
resonance.

The Gounaris-Sakurai (GS) parametrization~\cite{Gounaris:1968mw} is used to
describe the lineshape of the $\rho$ resonance decaying into two pions.
The parametrization takes the form
\begin{equation}
	R^{\rm
	GS}_{j}=\frac{1+\Gamma_{0}\cdot d/m_{0}}{m^{2}_{0}-m^{2}+f(m)-im_{0}\Gamma(m)}, 
\end{equation}
where $\Gamma(m)$ is given by Eq.~(\ref{gamma-mass-dep}).  Expressions for $f(m)$,
in terms of $\Gamma_{0}$ and $m$, and the constant $d$ can be found in
Ref.~\cite{Gounaris:1968mw}. 
The parameters specifying the $\rho$ lineshape are taken from
Refs.~\cite{Akhmetshin:2001ig,Barate:1997hv}, which provides lineshape information derived
from fits to \epem\ annihilation and $\tau$ lepton decay data. 

For the $J^{P}=0^{+}$ component of the $K\pi$ spectrum, denoted
$\left(K\pi\right)_{0}^{*0/+}$, we make use of the LASS
parametrization~\cite{Aston:1987ir}, which consists of a $K^{*}_{0}$
resonant term together with an effective-range, nonresonant component to
describe the slowly increasing phase as a function of the $K\pi$ mass:
\begin{eqnarray}
	R^{\rm
	LASS}_{j}=\frac{m}{\left|\vec{q}\right|\cot\delta_{\B}-i\left|\vec{q}\right|}+
	\nonumber \\
	e^{2i\delta_{\B}}\frac{m_{0}\Gamma_{0}\frac{m_{0}}{\left|\vec{q}_{0}\right|}}{\left(m_{0}^{2}-m^{2}\right)-im_{0}\Gamma_{0}\frac{\left|\vec{q}\right|m_{0}}{m\left|\vec{q}_{0}\right|}},
\end{eqnarray}
where
$\cot\delta_{\B}=\frac{1}{a\left|\vec{q}\right|}+\frac{1}{2}r\left|\vec{q}\right|$. The values used for the
scattering length $a$ and the effective range $r$ are given
in \tabref{signal-model-description}. The effective-range component 
has a cutoff imposed at $1800\mevcc$~\cite{Aubert:2008bj}. Integrating
separately the resonant term, the effective-range term, and the coherent
sum, we find that the \KstarII\ and the \KstarIIp\ resonances account for
$88\%$ of the sum, and the effective range component $49\%$; the $37\%$
excess is due to destructive interference between the two terms. The LASS
parametrization is the least-well-determined component of the signal model;
we discuss the impact of these uncertainties in \secref{systematics}.

The complex coefficients $c_{j}$ and $\bar{c}_{j}$ in
Eqs.~(\ref{isobar-formalism-A},\ref{isobar-formalism-Abar}) can be
parametrized in different ways; we follow the parametrization used in
Ref.~\cite{Aubert:2008bj} as it avoids a bias in the measurement of
amplitudes and phases when the resonant components have small magnitudes:
\begin{eqnarray}
	c_{j}       &=& \left(x_{j}+\Delta x_{j}\right)+i\left(y_{j}+\Delta y_{j}\right),\\
	\bar{c}_{j} &=& \left(x_{j}-\Delta x_{j}\right)+i\left(y_{j}-\Delta
	y_{j}\right), \nonumber
	\label{eq:complex-coeff}
\end{eqnarray}
where $x_{j}\pm\Delta x_{j}$ and $y_{j}\pm\Delta y_{j}$ are the real and
imaginary parts of the amplitudes. The quantities $\Delta x_{j}$ and $\Delta y_{j}$
parametrize the \CP\ violation in the decay. The \CP\ asymmetry for a given
intermediate state is given by
\begin{eqnarray}
	A_{\CP,j}&=&\frac{\left|\bar{c}_{j}\right|^{2}-\left|c_{j}\right|^{2}}{\left|\bar{c}_{j}\right|^{2}+\left|c_{j}\right|^{2}} \\ 
		 &=&-\frac{2\left(x_{j}\Delta x_{j}+y_{j}\Delta y_{j}\right)}{x_{j}^{2}+\Delta x_{j}^{2}+y_{j}^{2}+\Delta y_{j}^{2}}.
	\label{Acp-def}
\end{eqnarray}

The results quoted for the resonances in the following analysis use fit
fractions (${\rm F}{\rm F}_{j}$) as phase-convention-independent quantities
representing the fractional rate of each contribution in the Dalitz plot.
The ${\rm F}{\rm F}$ for mode $j$ is defined as
\begin{equation}
	{\rm F}{\rm F}_{j}=\frac{\int\int\left(\left|c_{j}F_{j}\right|^{2}+\left|\bar{c}_{j}\bar{F}_{j}\right|^{2}\right)dm^{2}_{K\pi}dm^{2}_{\pi\pi}}{\int\int\left(\left|A\right|^{2}+\left|\bar{A}\right|^{2}\right)dm^{2}_{K\pi}dm^{2}_{\pi\pi}}.
\end{equation}
The sum of all the fit fractions does not necessarily yield unity due
to constructive and destructive interference, as quantified by the
interference fit fractions given by~\cite{Bevan:2014iga}
\begin{equation}
	{\rm
FF}_{ij}=\frac{\int\int2Re\left[c_{i}c^{*}_{j}F_{i}F^{*}_{j}\right]dm^{2}_{K\pi}dm^{2}_{\pi\pi}}{\int\int\left|\sum_{k}c_{k}F_{k}\right|^{2}dm^{2}_{K\pi}dm^{2}_{\pi\pi}}.
\end{equation}

The parameters $x_{j}$, $\Delta x_{j}$, $y_{j}$, and $\Delta y_{j}$ are
determined in the fit, except for the reference amplitude. Fit fractions,
relative phases, and asymmetries are derived from the fit parameters and
their statistical uncertainties determined from pseudo experiments generated
from the fit results.

\section{The \babar\ detector and MC simulation}
\label{sec:detector}

The data used in the analysis were collected with the \babar\ detector at
the \pep2\ asymmetric-energy \epem\ collider at SLAC National Accelerator
Laboratory. The sample consists of \onreslumi\ of integrated luminosity
recorded at the \FourS\ resonance mass (``on-peak'') and \offreslumi\
collected 40\,\mev\ below the resonance mass
(``off-peak'')~\cite{Lees:2013rw}.  The on-peak sample corresponds to the
full \babar\ \FourS\ dataset and contains \bbpairs\ \BB\
events~\cite{Bevan:2014iga}. A detailed description of the \babar\ detector
is given in Refs.~\cite{Aubert:2001tu,TheBABAR:2013jta}. Charged-particle
tracks are measured by means of a five-layer double-sided silicon vertex
tracker (SVT) and a 40-layer drift chamber (DCH),  both positioned within a
solenoid that provides a 1.5~T magnetic field. Charged-particle
identification is achieved by combining the information from a ring-imaging
Cherenkov detector (DIRC) and specific ionization energy loss (\dedx)
measurements from the DCH and SVT. Photons are detected and their energies
measured in a CsI(Tl) electromagnetic calorimeter (EMC). Muon candidates
are identified in the instrumented flux return of the solenoid. 

We use \geant-based software to simulate the detector response and account
for the varying beam and experimental
conditions~\cite{Agostinelli:2002hh,Allison:2006ve}. The
\evtgen~\cite{Ryd:2005zz} and \jetset~\cite{Sjostrand:1995iq} software
packages are used to generate signal and background Monte-Carlo (MC) event
samples in order to determine efficiencies and evaluate background
contributions for different selection criteria.

\section{Event selection}
\label{sec:event-reco}

We reconstruct \BptoKspippiz\ candidates from one \piz\ candidate, one \KS\
candidate reconstructed from a pair of oppositely charged pions, and a
charged pion candidate. 
The \piz\ candidate is formed from a pair of neutral energy clusters in the
EMC with laboratory energies above $0.05~\gev$ and lateral
moments~\cite{Drescher:1984rt} between $0.01$ and $0.6$.  We require the
invariant mass of the reconstructed \piz\ to lie in the range $0.11<
m_{\g\g} <0.16~\gevcc$. The \KS\ candidate is required to have a $\pip\pim$
invariant mass within $15~\mevcc$ of the \KS\ mass~\cite{Beringer:1900zz},
and a proper decay time greater than $0.5\times10^{-11}{\rm s}$. To reduce
combinatorial background, we also require that the \KS\ candidates have a
vertex probability greater than $10^{-6}$ and that the cosine of the angle
between the \KS\ momentum direction and the \KS\ flight direction (as
determined by the interaction point and the \KS\ vertex) be greater than
$0.995$.  For the \pip\ candidate, we use information from the tracking
systems, the EMC, and the DIRC to select a charged track consistent with
the pion hypothesis.  We constrain the \pip\ track and \KS\ candidate to
originate from a common vertex. 

Signal events that are misreconstructed with the decay products of one or more daughters
completely or partially exchanged with other particles in the rest of the
event have degraded kinematic resolution. We refer to these as ``self-cross-feed''
(SCF) events. This
misreconstruction has a strong dependence on the energy of the particles
concerned and is more frequent for low-energy particles, \ie, for decays in
the corners of the Dalitz plot.  Because of the
presence of a \piz\ in the final state, there is a significant
probability for signal events to be misreconstructed due to low-energy
photons from the \piz\ decay.  Using a classification based on MC
information, we find that in simulated events the SCF fraction depends
strongly on the resonant substructure of the signal and ranges from $34\%$
for $\BptoKstarIppiz$ to $50\%$ for \BptorhopKs.  In events simulated
uniformly in phase space, hereafter referred to as nonresonant MC,
the SCF fraction varies from less than $10\%$ in the center of the Dalitz
plot to almost $70\%$ in the two corners of the Dalitz plot, where either
the \piz\ or the \pip\ has low energy. We describe how the SCF events are
handled in \secref{max-like}.

In order to suppress the dominant background, due to continuum
$\epem\to\qqbar\ (\q=u,d,s,c)$ events, we employ a boosted decision tree
(BDT) algorithm that combines four variables commonly used to discriminate jet-like
\qqbar\ events from the more spherical \BB\ events in the \epem\
center-of-mass (CM) frame.  The first of these is the ratio of the
second-to-zeroth order momentum-weighted Legendre polynomial moments,
\begin{equation}
\frac{L_2}{L_0} =
\frac
{\sum\limits_{i\in{\rm ROE}}\frac{1}{2}\left(3\cos^{2}{\theta_{i}}-1\right)p_{i}}
{\sum\limits_{i\in{\rm ROE}}p_{i}},
\label{eq:Legendre}
\end{equation}
where the summations are over all tracks and neutral clusters in the event,
excluding those that form the \B\ candidate (the ``rest of the event'' or ROE);
$p_{i}$ is the particle momentum, and $\theta_{i}$ is the angle between the
particle and the thrust axis of the \B\ candidate, hereafter also referred
to as the \B.
The three other variables entering the BDT are the absolute value of
the cosine of the angle between the \B\ direction and the collision axis,
the zeroth-order momentum-weighted Legendre polynomial moment, and the
absolute value of the output of another BDT used for ``flavor tagging'',
\ie, for distinguishing \B\ from \Bb\ decays using inclusive properties of
the decay of the other \B\ meson in the $\FourS\to\B\Bbar$
event~\cite{Aubert:2009yr}.  The momentum-weighted Legendre polynomial
moments and the cosine of the angle between the \B\ direction and the beam
axis are calculated in the \epem\ CM frame.  The BDT is
trained on a sample of signal MC events and off-peak data.  We apply a
loose criterion on the BDT output of $\NN>0.06$, which retains
approximately $70\%$ of the signal while rejecting $92\%$ of the $\qqbar$
background.

In addition to \NN, we use two kinematic variables to distinguish the
signal from the background:
\begin{eqnarray}
\mes &=& \sqrt{E_{\rm X}^{2}-{\bf p}^{2}_{\B}} \,, \\
\DeltaE &=& E^{\star}_{\B} - \sqrt{s}/2 \,,
\end{eqnarray}
where
\begin{equation}
	E_{\rm X} = \left(s/2 + {\bf p}_{\epem} \cdot {\bf
	p}_{\B}\right)/E_{\epem} \,,
\end{equation}
and where $\sqrt{s}$ is the total \epem\ CM energy, with
$\left(E_{\epem},{\bf p}_{\epem}\right)$ and $\left(E_{\B},{\bf p}_{\B}\right)$ the
four-momenta of the initial \epem\ system and the \B\ candidate, respectively, both
measured in the lab frame, while the star indicates the \epem\ CM frame. The
signal \mes\ distribution for correctly reconstructed events is
approximately independent of their position in the
$\KS\pip\piz$ Dalitz plot and peaks near the \B\ mass with a
resolution of about $3.4\mevcc$.  

We retain all candidates satisfying the following selection
criteria: $5.23<\mes<5.29\gevcc$ and $-0.3<\DeltaE<0.3\gev$. The
signal region, where the final fit to data is performed, is defined by the
tighter criteria
$5.260<\mes<5.287\gevcc$ and $-0.20<\DeltaE<0.15\gev$. We also use candidates
in the sideband region of \mes\ defined by
$5.23<\mes<5.26\gevcc$ and $-0.20<\DeltaE<0.15\gev$ and subtract from
distributions for these
events the \BB\ background contributions predicted by MC simulations. We
then add these distributions to the off-peak data distributions to increase the
statistical precision of our model of the Dalitz plot distribution for
continuum background. 

Each of the \B\
candidates is refit to determine the Dalitz plot variables. In these fits
the $\KS\pip\piz$ invariant mass is constrained to the world average value of
the \B\ mass~\cite{Beringer:1900zz} to improve position resolution within the
Dalitz plot. 

We find that $20\%$ of the remaining events in nonresonant MC have two or more candidates. We
choose the best candidate in multiple-candidate events based on the highest
\B-vertex probability.  This procedure is found to select a correctly
reconstructed candidate more than $60\%$ of the time and does not bias the
fit variables.

The reconstruction efficiency over the Dalitz plot is modeled
using a two-dimensional (2D) binned distribution based on a generated sample of
approximately $2\times10^{6}$ simulated \BptoKspippiz\ MC events, where the events
uniformly populate phase space. All selection criteria are applied except
for those corresponding to a $K\pi$ invariant-mass veto described below,
which is taken into account separately. The 2D histogram of reconstructed
MC events is then divided by the 2D histogram of the generated MC events.
In order to expand regions of phase space with large efficiency variations,
the Dalitz plot variables are transformed into ``square Dalitz
plot''~\cite{Aubert:2007jn} coordinates.  We obtain an average efficiency,
for nonresonant MC events, of approximately $15\%$. In the likelihood fit
we use an event-by-event efficiency that depends on the Dalitz plot position.

\section{\BB\ backgrounds}
\label{sec:bkg-study}

In addition to continuum events, background arises from non-signal \BB\
events.  A major source of \BB\ background arises from
$\Bp\to\Dzb\left(\to\KS\piz\right)\pip$ decays. To suppress this
background, we veto events with $1.804<m_{\KS\piz}<1.924\gevcc$. 

The remaining \BB\ backgrounds are studied using MC simulations and classified based
on the shape of the \mes, \DeltaE, and Dalitz plot distributions. We identify nine
categories of \BB\ backgrounds: 
categories 1, 2 and 3 include different types of three- and four-body \B\
decays involving an intermediate \D\ meson; 
categories 4 and 5 include charmless four-body \B\ decays to intermediate
resonances where a \piz\ in the final state is not reconstructed;
categories 6 and 7 include two-body \B\ decays with a radiated photon
misreconstructed as a \piz\ decay product or where the \piz\ arises from
the other \B\ decay; 
category 8 includes charmless three-body \B\ decays where a charged pion is
interchanged with a \piz\ meson from the other \B; 
and finally category 9 includes all other simulated \BB\ background
contributions. 
Within each category, each of the \mes, \DeltaE, \NN, and Dalitz plot
distributions are formed by combining the contributions of all decay modes
in the category. The combinations are done by normalizing the distributions
for each decay mode to the expected number of events in the recorded data
sample, which is estimated using reconstruction efficiencies determined
from MC, the number of \BB\ pairs in the recorded data sample, and the
branching fractions listed in Refs.~\cite{Beringer:1900zz,Amhis:2012bh}.
For each category, the histograms of \mes, \DeltaE, \NN, and the Dalitz
plot variables are used as the probability density functions (PDF) in the
likelihood fit to data to model the \BB\ background. 

\section{The maximum likelihood fit}
\label{sec:max-like}

The extended likelihood function is given by
\small
\begin{eqnarray}
	&&{\cal L}=\exp\left(-\sum_{k}N_{k}\right)\times \\ \nonumber
	&& \prod^{N_{e}}_{i=1}\left[\sum_{k}N_{k}{\cal
	P}_{k}^{i}\left(\mkspipSq,\mpippizSq,\mes,\DeltaE,\NN,q_{\B}\right)\right]
	\,,
\end{eqnarray}
\normalsize
where $N_{k}$ is the number of candidates in each signal or background
category $k$, $N_{e}$ is the total
number of events in the data sample, and ${\cal P}_{k}^{i}$ (the PDF for
category $k$ and event $i$) is the product of the PDFs
describing the Dalitz plot, \mes, \DeltaE, and \NN\ distributions, with
$q_{\B}$ the charge of the \B\ candidate.  

To avoid possible biases in the determination of the fit
parameters~\cite{Punzi:2004wh}, we use MC samples to study correlations
between the fit variables and the Dalitz plot parameters, \mkspipSq\ and
\mpippizSq. We find that for correctly reconstructed signal candidates, the
\DeltaE\ distribution is strongly dependent on \mkspip. This is mostly due
to a dependence of the energy resolution of the \B\ candidate on the \piz\
momentum. For SCF signal candidates, both the \mes\ and \DeltaE\
distributions depend on all three two-body invariant masses: \mkspip,
\mkspiz, and \mpippiz. The \mes, \DeltaE, and \NN\ distributions for
continuum and \BB\ backgrounds have negligible correlations with the Dalitz
plot parameters.

\begin{figure}
	\includegraphics[width=0.5\textwidth]{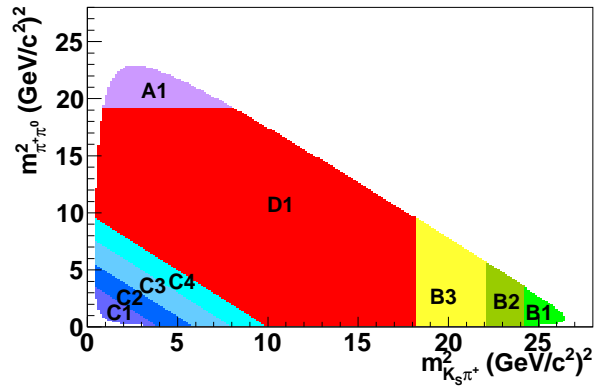}
	\caption{Diagram illustrating the division of the Dalitz plot into
		different regions for the definition of the PDFs for
		self-crossfeed signal events.  Each letter indicates
		whether the dependence is on $\mpippizSq$ (A), $\mkspipSq$
		(B), or $\mkspizSq$ (C). The remaining region of the Dalitz
		plot (D1) is where we expect to find fewer SCF events, and
		where the shapes for \mes\ and \DeltaE\ are less dependent
		on their position in the Dalitz plot, further described in
		\tabref{scf-pdf-list}.  
	} 
\label{fig:scf-map} 
\end{figure}

For correctly reconstructed signal candidates, the \mes\ and \DeltaE\ PDFs
are parameterized by a Cruijff function, which is given by (omitting
normalization factor)
\begin{equation}
	f_{\rm Cruijff}(x)=\exp\left[\frac{-\left(x-m\right)^{2}}{2\sigma_{L,R}^{2}+\alpha_{L,R}\left(x-m\right)^{2}}\right],
	\label{eq:Cruijff-func}
\end{equation}
where $m$ gives the peak of the distribution and the asymmetric width of
the distribution is given by $\sigma_{L}$ for $x<m$ and $\sigma_{R}$ for
$x>m$. The asymmetric modulation is similarly given by $\alpha_{L}$ for
$x<m$ and $\alpha_{R}$ for $x>m$.
The \DeltaE\ PDF parameters are calculated on an event-by-event basis in
terms of the $\KS\pip$ invariant mass, as a linear function for
$\mkspipSq<20\gevccSq$ and as a quadratic function for
$\mkspipSq>20\gevccSq$.  These functions are determined by fitting the
\DeltaE\ distribution in large nonresonant MC samples.  For the SCF signal,
in order to follow the rapid shape variations across the Dalitz plot of the
\mes\ and \DeltaE\ distributions, we divide the Dalitz plot into several
regions as illustrated in \figref{scf-map}. Each letter indicates whether
the dependence is on \mpippizSq, \mkspipSq, or \mkspizSq. The regions are
chosen based on the distribution in the Dalitz plot of the SCF fraction and
the mean difference between the true and reconstructed position in the
Dalitz plot; we include more regions in areas of the Dalitz plot where
these quantities are largest.  We use \mes\ and \DeltaE\
PDFs specific to each region, as listed in \tabref{scf-pdf-list}. Some
of the PDFs used in the parametrization of the SCF include Cruijff
functions, Chebychev polynomials, Gaussian functions, and two-piece Gaussian
(BGauss) functions. A two-piece Gaussian function is an asymmetric Gaussian described by the
following functional form (omitting normalization factor)
\begin{equation}
	f_{\rm BGauss}(x)=\exp\left[\frac{-\left(x-m\right)^{2}}{2\sigma_{L,R}^{2}}\right].
	\label{eq:BGauss-func}
\end{equation}

\begin{table}
	\caption{
		List of PDFs used to describe the \mes\ and \DeltaE\
		self-crossfeed signal distributions in each of the regions
		of the \BptoKspippiz\ Dalitz plot shown in
		\figref{scf-map}. The abbreviations correspond to the
		following functional forms: Cruijff function described in
		Eq.~(\ref{eq:Cruijff-func}) (Cruijff), Chebychev polynomial
		(Cheb), Gaussian (Gauss), two-piece Gaussian described in
		Eq.~(\ref{eq:BGauss-func}) (BGauss), and exponential (Exp).
	}
	\label{tab:scf-pdf-list}
	\begin{tabular}{r|c|c}
		\hline
		Dalitz plot region & \mes\ PDF	& \DeltaE\ PDF	\\
		\hline
\phantom{\huge I}$\mpippizSq$ (A1) & Cruijff	& Cruijff	\\
		$\mkspipSq$ (B1)  & Cheb+Gauss	& Exp+Sigmoid	\\
			      (B2)  & Cheb+Gauss	& linear+BGauss	\\
			      (B3)  & Cruijff	& Exp+Sigmoid	\\
		$\mkspizSq$ (C1)  & Cheb+Gauss	& Cheb 		\\
			      (C2)  & Cheb+Gauss	& Cheb		\\
			      (C3)  & Cruijff	& Cheb		\\
			      (C4)  & Cruijff	& Cruijff	\\
		Central region (D1) & Cruijff	& Cruijff	\\
		\hline
	\end{tabular}
\end{table}

For the continuum background, we use an ARGUS function~\cite{Albrecht:1990am}
to parameterize the \mes\ shape.  The \DeltaE\ distribution 
is described by a linear function, and the \NN\ distribution by an
exponential function.  The \mes, \DeltaE, and \NN\ PDFs for \BB\ backgrounds are
defined by the sum of the histograms from the MC simulations for
decay modes in each background category, as described in \secref{bkg-study}.  

The continuum and \BB\ background Dalitz plot distributions are
included in the likelihood as two-dimensional histograms. For \BB\
backgrounds, we use MC samples. For continuum background, we
combine events from the off-peak data and the \mes\ sideband in
on-peak data, after subtracting contributions from \B\ decays, as
described in \secref{event-reco}.  For the 2D histograms, we
use the square Dalitz plot coordinates. A linear interpolation between bin
centers is applied.

The free parameters in the fit are the yields for signal, continuum
background, and \BB\ background categories 1 and 9.  The yields for the
remaining \BB\ background categories are fixed to the estimated values.
All the PDF parameters for the correctly reconstructed \mes\ and \DeltaE\ PDFs, except for the
tail parameters, are determined in the fit. All SCF signal PDF parameters are
fixed to values obtained from fits to nonresonant MC events.  The endpoint of
the ARGUS function is fixed to $5.289\gevcc$ while the shape parameter is
determined in the fit.  The slope for the linear function of the \DeltaE\
PDF and the exponent for the exponential function of the \NN\ PDF for
continuum background are similarly determined in the fit.  The isobar
coefficients, $x$ and $y$ in Eq.~(\ref{eq:complex-coeff}), for all but one of
the isobar components are fitted parameters in the fit and are measured
relative to the fixed isobar component. The coefficients for the reference
isobar are fixed to $x=1$ and $y=0$. In total, the fit is performed with
$21$ free parameters.

\begin{table}[h]
	\caption{Fit fractions obtained from the fit to data when
		each additional isobar is added to the fit model 
		one at a time.
	}
	\label{tab:addition-res-fitfrac}
	\begin{tabular}{l|c}
		\hline
		Additional isobar &	Fit fraction 	\\ 
		\hline
\phantom{\huge I}\rhoIIp	& $0.042\pm 0.044$	\\
		\KstarIII	& $0.038\pm 0.017$	\\
		\KstarIIIp	& $0.012\pm 0.020$	\\
		\KstarIV	& $0.032\pm 0.034$	\\
		\KstarIVp	& $0.005\pm 0.030$	\\
		\hline
	\end{tabular}
\end{table}

We determine a nominal signal Dalitz plot model based on information from
previous
studies~\cite{Aubert:2008bj,BABAR:2011aaa,Aubert:2009me,BABAR:2011ae}, and
on the changes in the log likelihood in the fit to data when resonances are
added to, or
removed from, the list shown in \tabref{signal-model-description}.
In these fits to the combined \Bp\ and \Bm\ data samples, the \CP\
coefficients $\Delta x$ and $\Delta y$ are fixed to zero. We do not find
significant contributions in the fit when adding the resonances \rhoIIp,
\KstarIII, \KstarIIIp, \KstarIV, or \KstarIVp, one at a time to the
default model. We observe that the fit fractions for these additional
resonances, reported in \tabref{addition-res-fitfrac}, are consistent with
zero. The most statistically significant of these fit fractions is ${\rm
FF}\left(\KstarIII\pip\right)=0.038\pm0.017$; since the statistical
significance is less than $2$ standard deviations, we do not include any of
the additional resonances in the nominal fit.  

 \begin{figure*}
 \includegraphics[width=.45\textwidth]{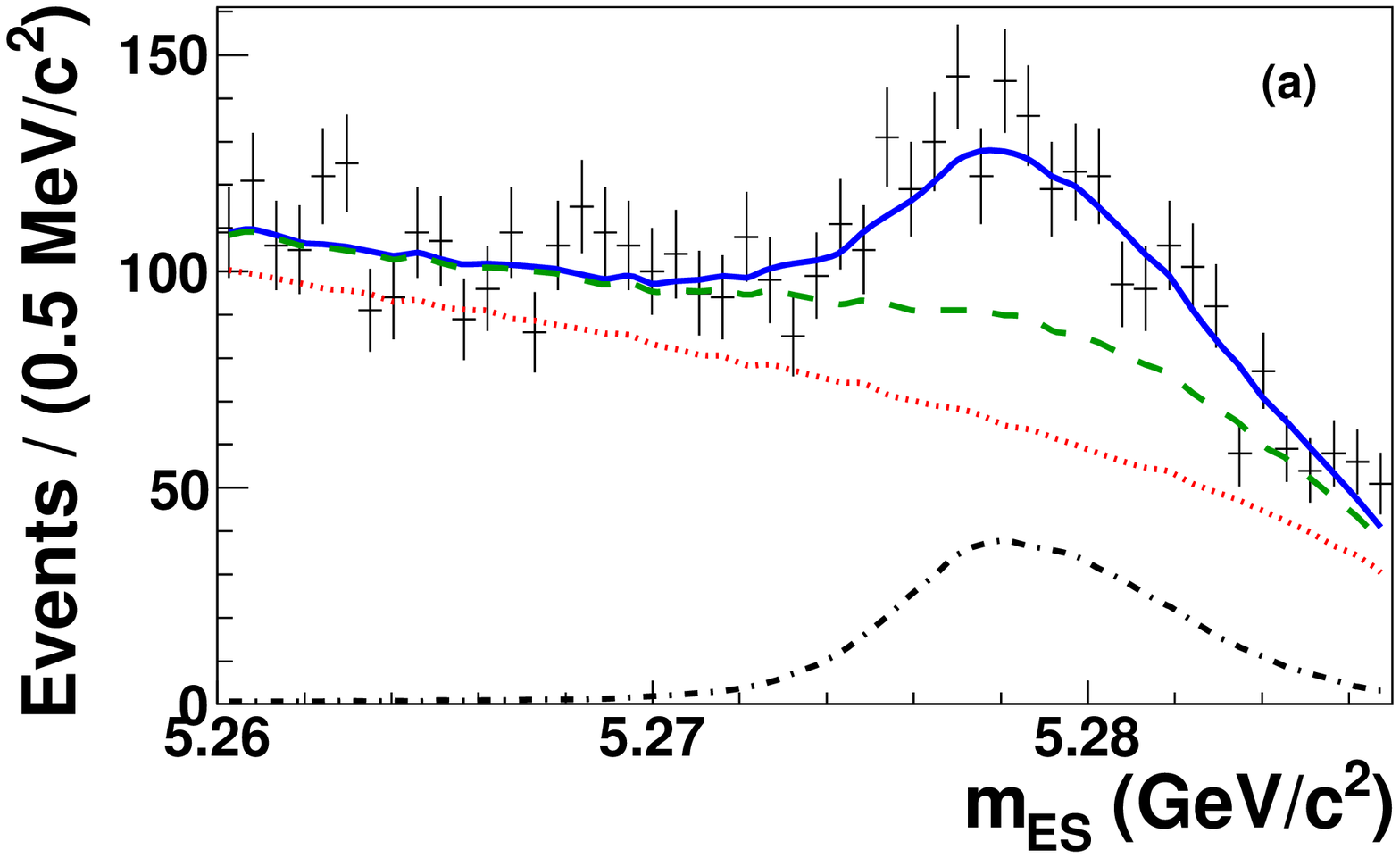} 
 \includegraphics[width=.45\textwidth]{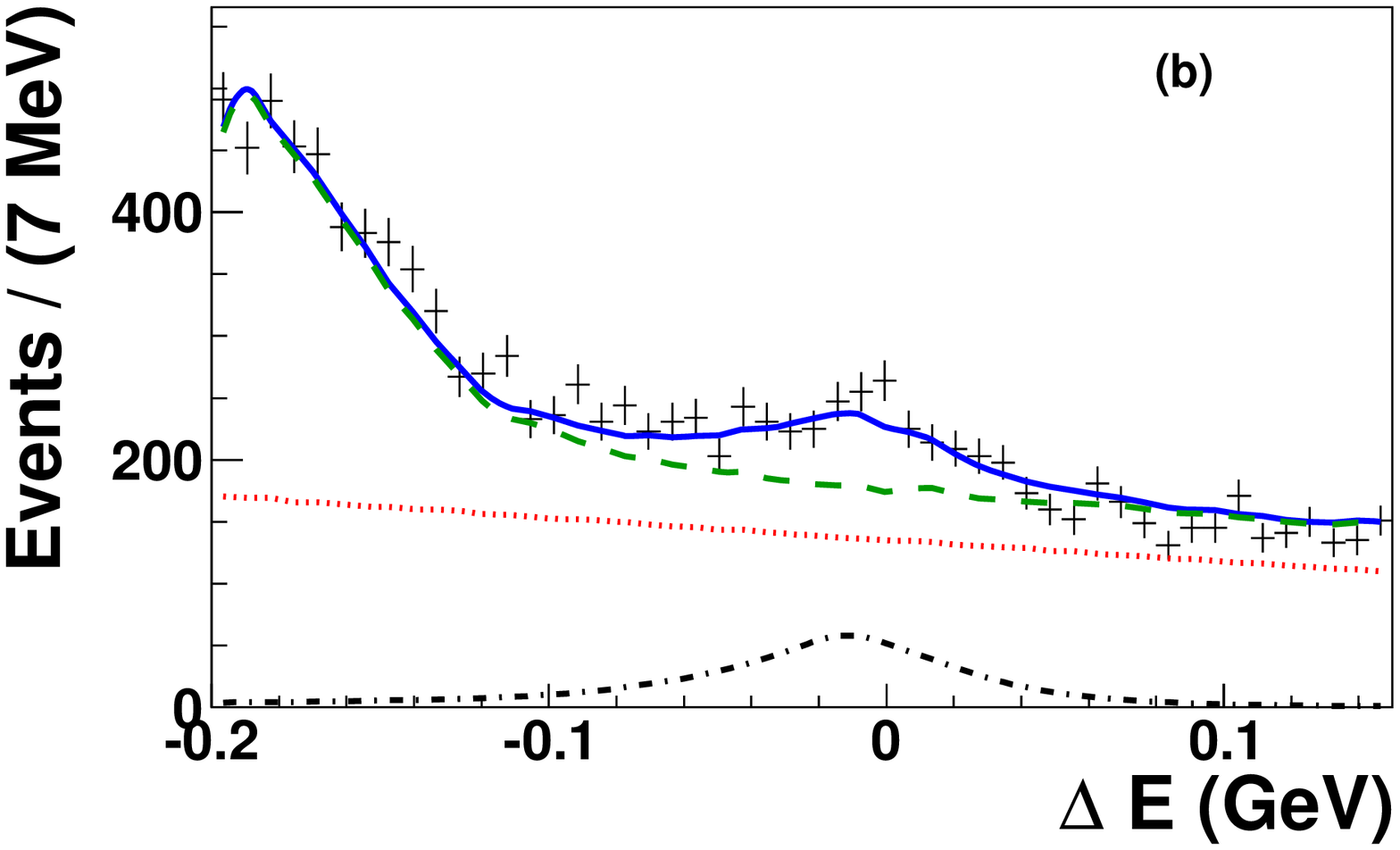}
 \includegraphics[width=.45\textwidth]{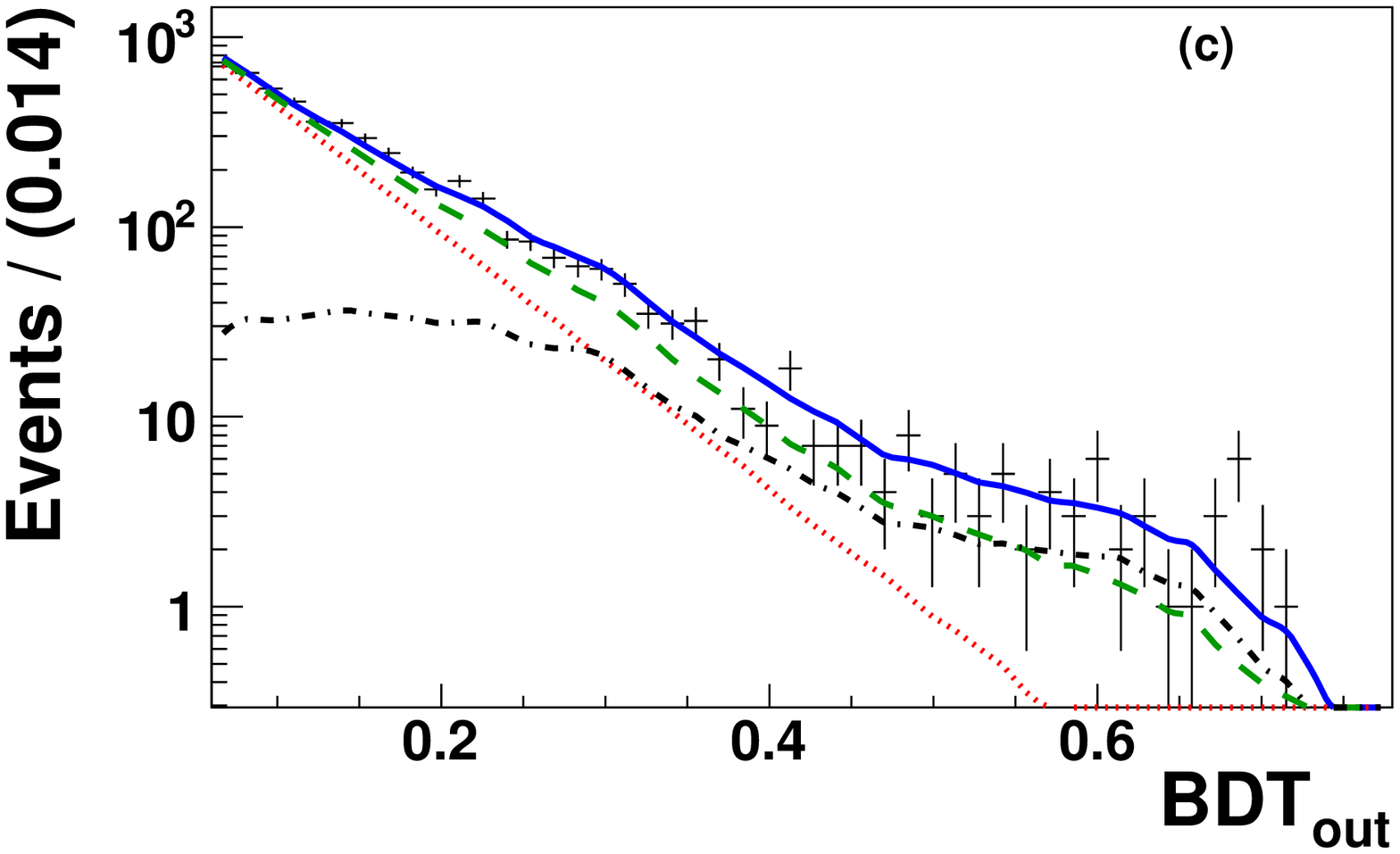}
 \includegraphics[width=.45\textwidth]{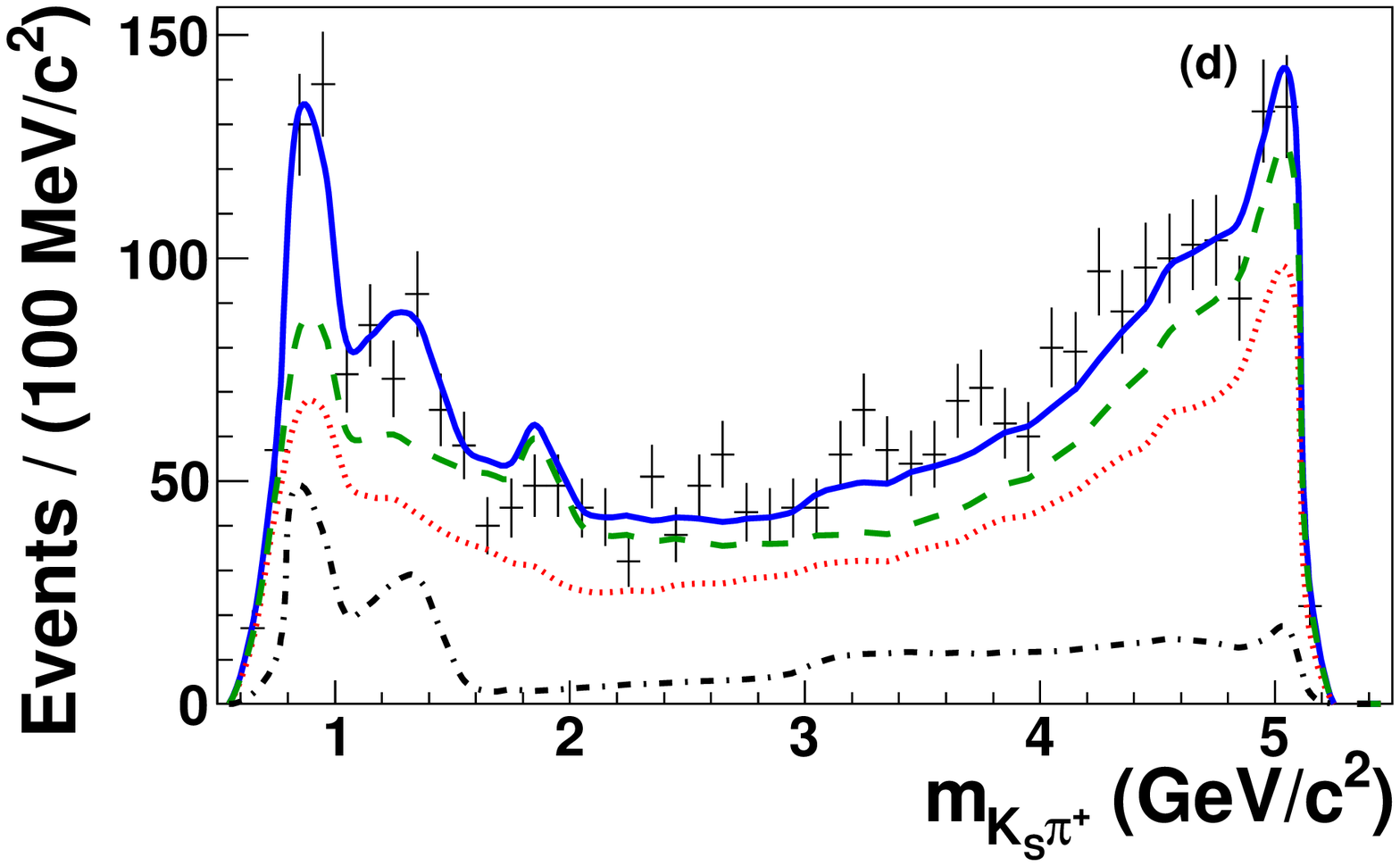} 
 \includegraphics[width=.45\textwidth]{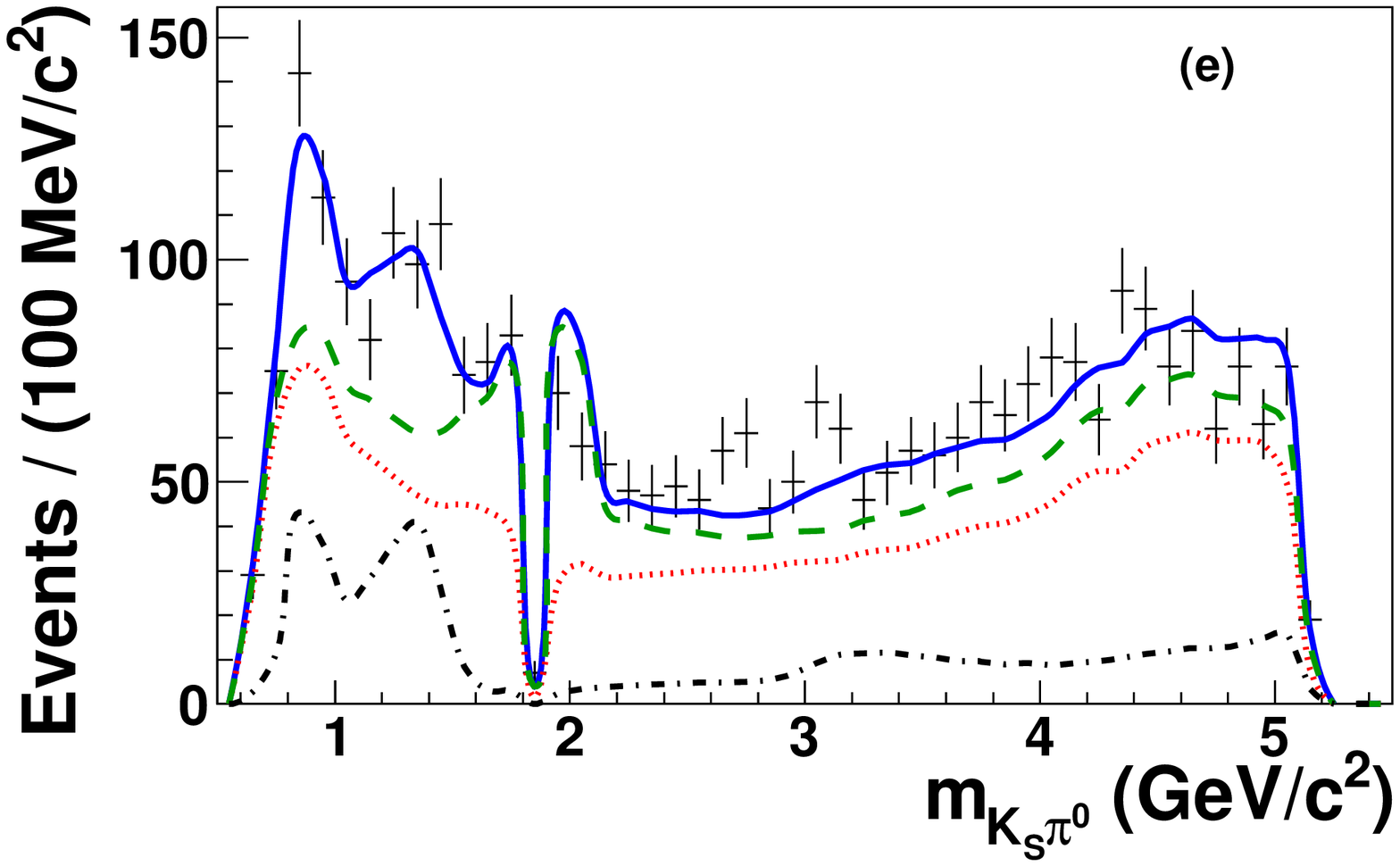}
 \includegraphics[width=.45\textwidth]{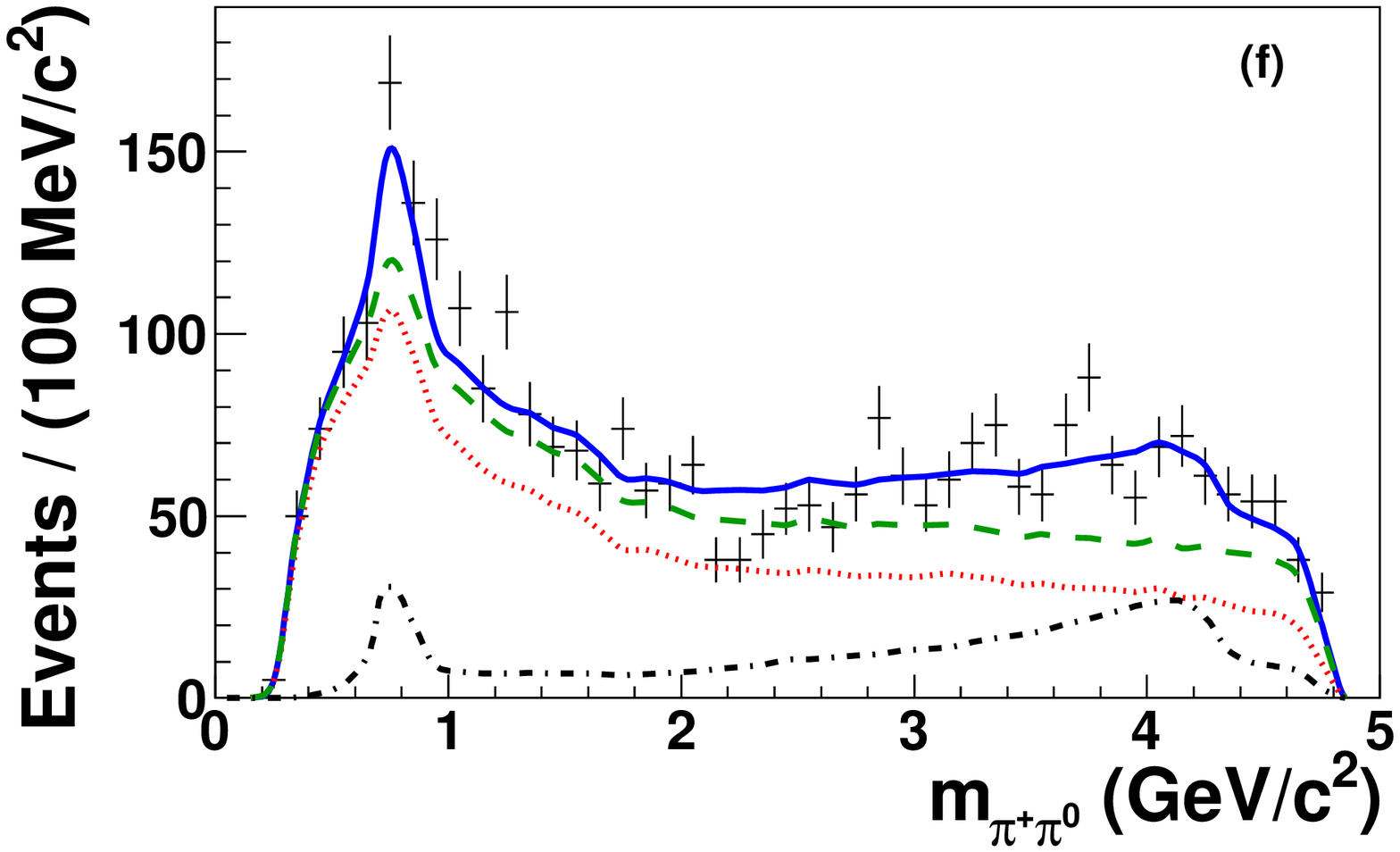}
 \caption{
	 {Combined \Bpm\ fit}: Measured distributions and fit projections
	 for $\Bpm\to\KS\pipm\piz$ candidates;
   (a) \mes, 
   (b) \DeltaE, 
   (c) \NN, 
   (d) \mkspip, 
   (e) \mkspiz, and
   (f) \mpippiz.
   The points with error bars correspond to data, the solid (blue) curves to the
   total fit result, the dashed (green) curves to the total background  
   contribution, and the dotted (red) curves to the continuum background
   component. The dash-dotted curves represent the signal contribution.
   The projected distributions are obtained from statistically precise
   pseudo experiments generated using the fit results.  For all
   distributions in each panel, the signal-to-background ratio is increased
   by applying tighter selection requirement on \mes, \DeltaE, and/or \NN,
   listed \tabref{signal-enhance-selections}.
 }
 \label{fig:signal-project}
 \end{figure*}

We do not observe an excess of events for invariant masses greater than
$2\gevcc$, suggesting that a nonresonant component, in addition to that
included in the LASS parametrization, is not necessary.
We observe that if we add a nonresonant component to the fit, the 
change in log likelihood for the binned data and the fit projections
for the $\KS\pip$, $\KS\piz$, and $\pip\piz$ invariant masses are
consistent with the expected change due to the additional free parameters in
the fit, and do not indicate any statistically significant nonresonant
component. We therefore conclude that, with the current level of
statistical sensitivity, the base model, which includes the \rhoIp,
\KstarIp, \KstarI, \KpiSwave, and \KpiSwavep\ resonances, provides an
adequate description of the data.

\section{Results}
\label{sec:results}

We apply the fit described in \secref{max-like} to the \ncand\ selected 
$\Bp\to\KS\pip\piz$ candidates. A first fit is performed on the combined
\Bpm\ sample. We obtain yields of \nsig\ signal events, $24\,381\pm200$
continuum events, $2745\pm70$ \BB\ events in category 1, and $1768\pm140$
\BB\ events in category 9.  The results of the fit are shown in
Fig.~\ref{fig:signal-project}. For the purpose of this figure, the
contributions of signal events are enhanced by applying the more
restrictive selection criteria listed in
\tabref{signal-enhance-selections}.

\begin{table}[h]
	\caption{
		Selection criteria imposed to enhance the contributions of
		signal events for the results presented in
		Figs.~\ref{fig:signal-project} and
		\ref{fig:signal-project-cpFit}.
	}
	\label{tab:signal-enhance-selections}
	\begin{tabular}{l|l}
		\hline
		Projection plot	& Selections	\\
		\hline
		\mes	& $-0.05<\DeltaE<0.05\gev$	\\
			& $\NN>0.1$	\\
		\hline
		\DeltaE	& $\mes>5.27\gevcc$	\\
			& $\NN>0.1$	\\
		\hline
		\NN	& $\mes>5.27\gevcc$		\\
			& $-0.05<\DeltaE<0.05\gev$	\\
		\hline
		$m_{\KS\pip}$, $m_{\KS\piz}$, $m_{\pip\piz}$ &
			$\mes>5.27\gevcc$ \\
			& $-0.05<\DeltaE<0.05\gev$	\\
			& $\NN>0.1$	\\
		\hline
	\end{tabular}
\end{table}

\begin{table*} 
	\caption{{Combined \Bpm\ fit}: Relative phases, $\phi$, for the isobar amplitudes as
		measured from five fits to data, where each of the five
		isobar amplitudes is in turn taken as the reference.  All
		phases are quoted in degrees.  The uncertainties are
		statistical only.  
	}
	\label{tab:onpeak-phases}
	\begin{tabular}{l|ccccc}
		\hline
		& \\
		& \multicolumn{5}{c}{Relative phase (degrees)}	\\
\backslashbox{Reference amplitude}{Resonant contribution} & $\KstarI\pip$ &
				 $\KstarIp\piz$ & $\KpiSwave\pip$ &
				 $\KpiSwavep\piz$ & $\rhoIp\KS$\\
		\hline
		\BptoKstarIpip\phantom{\large I}	& \phantom{-00}$0\phantom{0}$     & $-95\pm43$  & \phantom{-}$174\pm11$ & $-89\pm43$	    & $-122\pm43$	\\
		\BptoKstarIppiz  	& -- 		& \phantom{-0}$0$    & \phantom{0}$-90\pm42$ & \phantom{-0}$6\pm10$  & \phantom{0}$-27\pm26$	\\
		\BptoKpiSwavepip 	& --			   & --			& \phantom{-00}$0$  & \phantom{-}$96\pm42$  & \phantom{-0}$63\pm37$	\\ 
		\BptoKpiSwaveppiz	& --			   & --			& --			& \phantom{-0}$0$   & \phantom{0}$-32\pm25$	\\ 
		 \BptorhopKs    	& --			   & --			& --			& --		& \phantom{-00}$0$	\\ 
		\hline
	\end{tabular}
\end{table*}

\begin{table*}
	\caption{ {Combined \Bpm\ fit}: 
		Results for the fit fractions ${\rm FF}_{j}$ (diagonal
		terms) and interference terms ${\rm FF}_{ij}$ in data for each resonant
		contribution.
		The uncertainties are statistical only. 
	}
	\label{tab:onpeak-fitfrac}
	\begin{tabular}{l|ccccc}
		\hline
		&
		\multicolumn{5}{c}{${\rm FF}_{j}$ and ${\rm FF}_{ij}$} \\
		\hline
		Resonant contribution	& $\KstarI\pip$	& $\KstarIp\piz$ & $\KpiSwave\pip$	& $\KpiSwavep\piz$	& $\rhoIp\KS$ \\
		\hline
	\BptoKstarIpip\phantom{\large I} & $0.10\pm0.03$ & $0.0004\pm0.0028$ & $(17\pm5)\times10^{-5}$ & $0.007\pm0.005$		   & $-0.008\pm0.007$ \\
	\BptoKstarIppiz	  	& --		& $0.14\pm0.02$     & $-0.010\pm0.007$	      & $(-3\pm1)\times10^{-6}$	   & $0.012\pm0.008$ \\
	\BptoKpiSwavepip	& --		& --		    & $0.36\pm0.05$ 	      & $(1.5\pm6.1)\times10^{-5}$ & $-0.04\pm0.02$	\\
	\BptoKpiSwaveppiz	& --		& --		    & --		      & $0.27\pm0.03$	           & $-0.02\pm0.02$\\
	\BptorhopKs		& --		& --                & --		      & --	                   & $0.19\pm0.04$	\\
		\hline
	\end{tabular}
\end{table*}

The branching fraction for \BptoKzpippiz\ is determined from the
number of signal events, the efficiency estimated from MC events, and the total
number of \BB\ events in data. We take into account differences
between the \piz\ reconstruction efficiency in data and MC events,
determined from control samples with either $\tau$ leptons or
initial-state radiation, as a function of \piz\ momentum
($\frac{\epsilon_{\rm data}}{\epsilon_{\rm MC}}=97.2\%$, averaged over
\piz\ momentum). We correct for small biases in the branching fraction, as
determined from MC pseudo experiments generated with the same number of
signal events and resonance composition as found in the fit to data.  We
divide the partial branching fraction of $\Bp\to\KS(\to\pip\pim)\pip\piz$
by the branching fraction for $\KS\to\pip\pim$, and multiply the result by a
factor of $2$ to account for \KL\ decay, to obtain the branching fraction
result ${\cal B}\left(\Bp\to\Kz\pip\piz\right) =
\left(\kpipiBFal\right)\times10^{-6}$, where the first uncertainty is
statistical, the second is systematic, and the third is due to assumptions
made concerning the signal model. The latter two uncertainties are
described in \secref{systematics} and the breakdown of the systematic
uncertainties is detailed in \tabref{systematics-bf}.

We measure amplitudes and phases relative to each of the five two-body
decays in the signal model to take advantage of the smaller uncertainty
observed when measuring the relative phases of the two pairs of decays with
same-charge $K^{*}$ resonances.  \tabref{onpeak-phases} lists the relative
phase, $\phi$, between each pair of two-body decays in the signal model and
its uncertainty.  The statistical uncertainty in the relative phase is
smallest $(\approx10^{\circ})$ for the resonances that decay to the
same-charge $K\pi$ state. This is due to a larger overlap in the Dalitz
plot between the same-charge \Kstar\ resonances than occurs for other pairs
of resonances that only overlap in the corners of the Dalitz plot. 

Since the statistical uncertainties of the fit fractions do
not depend on the reference mode, we quote in \tabref{onpeak-fitfrac} only
the fit fractions from a fit relative to the $\KstarI\pip$ amplitude. The
fit fractions for the $\KstarII\pip$ and $\KstarIIp\piz$ modes are the
product of the $(K\pi)^{*}_{0}$ S-wave fit fraction, shown in
\tabref{onpeak-fitfrac}, and the fraction due to
the resonant contribution in the LASS parametrisation ($88\%$).  The
off-diagonal fit fractions are small compared to the diagonal elements. We
calculate the branching fractions for the resonant contributions shown in
\tabref{results-bf-Acp} as the product of the total branching fraction
and the fit fractions returned by the fit to data, including appropriate
Clebsch-Gordan coefficients.

 \begin{figure*}[!htb]
 \includegraphics[width=.45\textwidth]{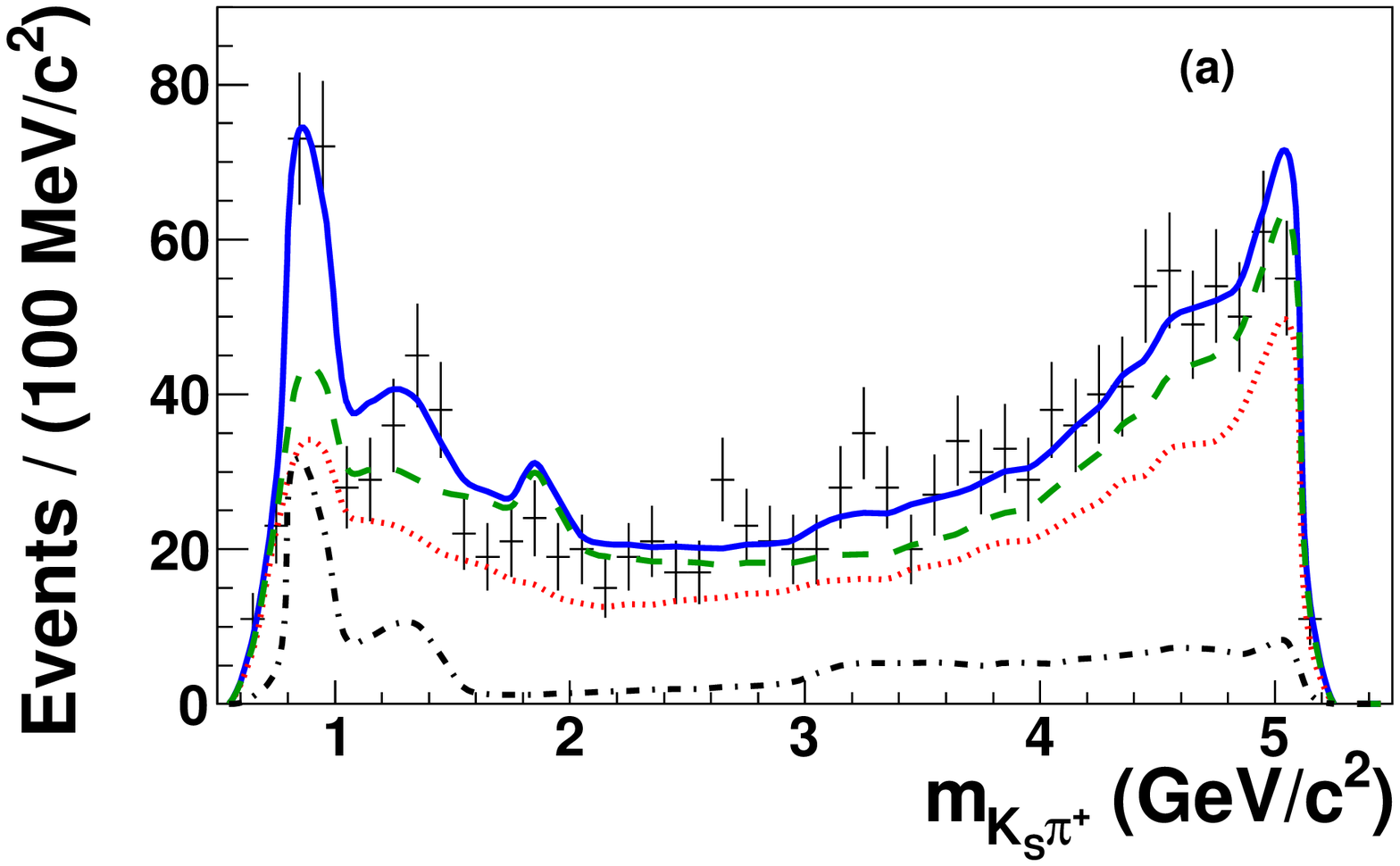} 
 \includegraphics[width=.45\textwidth]{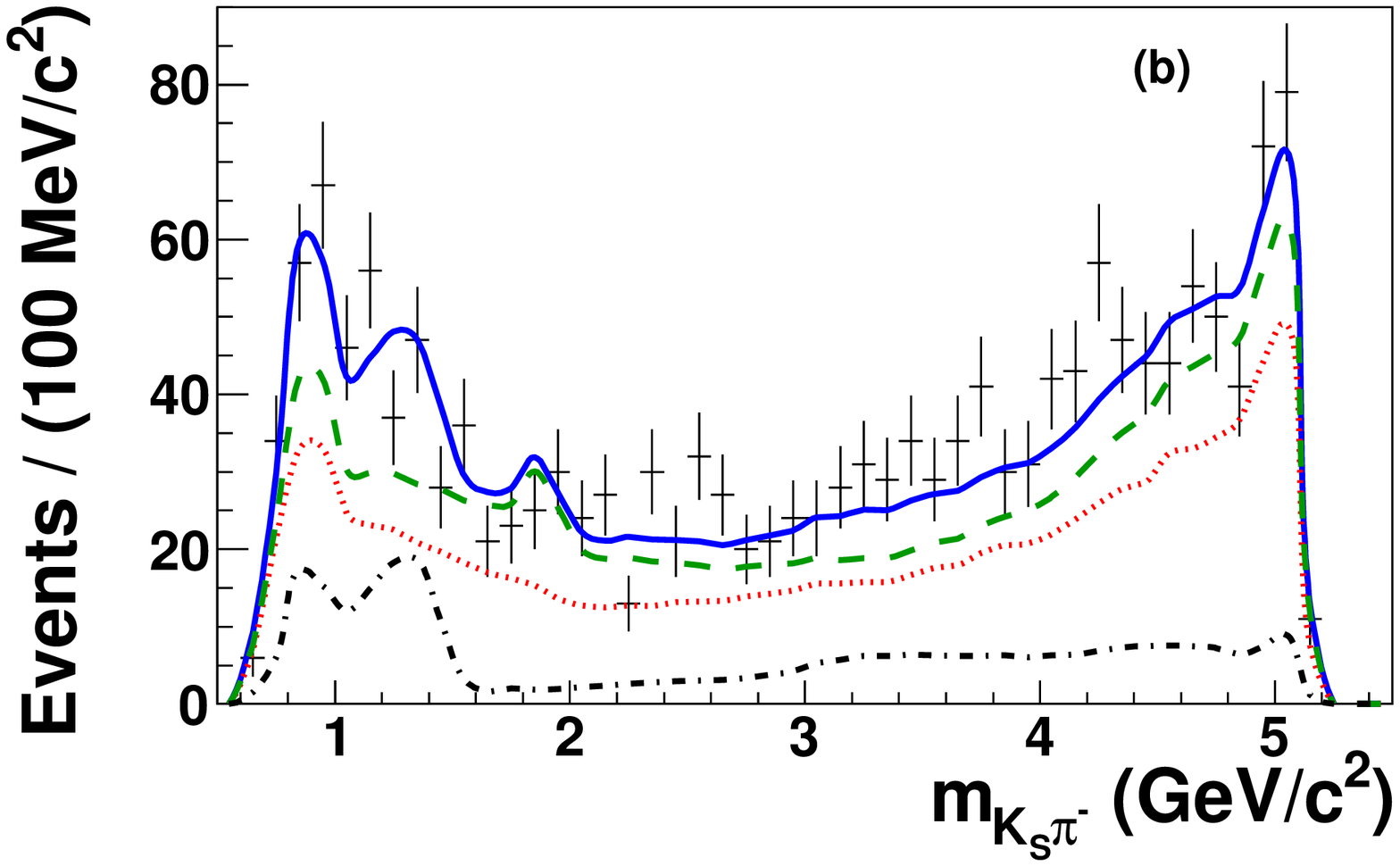} 
 \includegraphics[width=.45\textwidth]{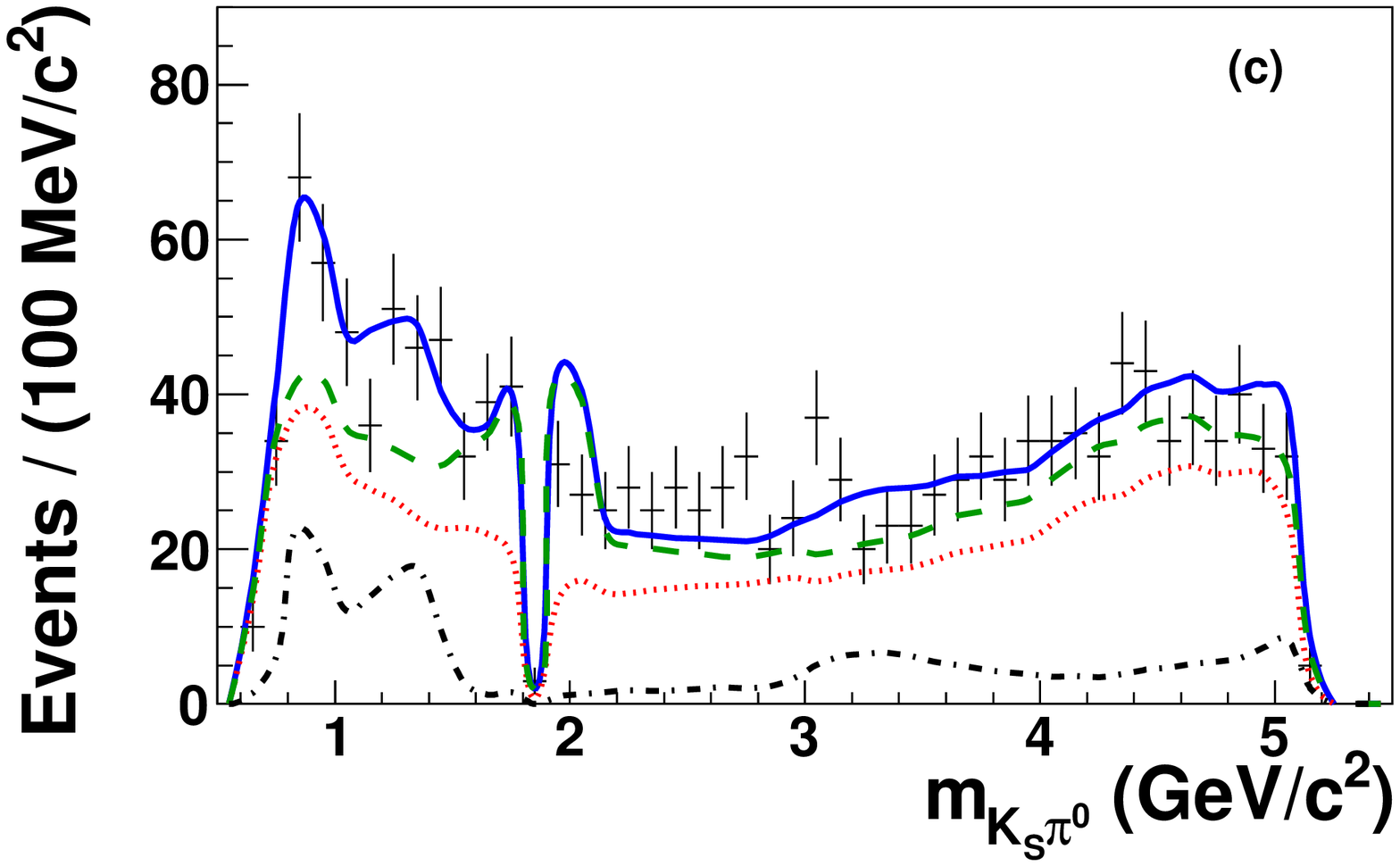}
 \includegraphics[width=.45\textwidth]{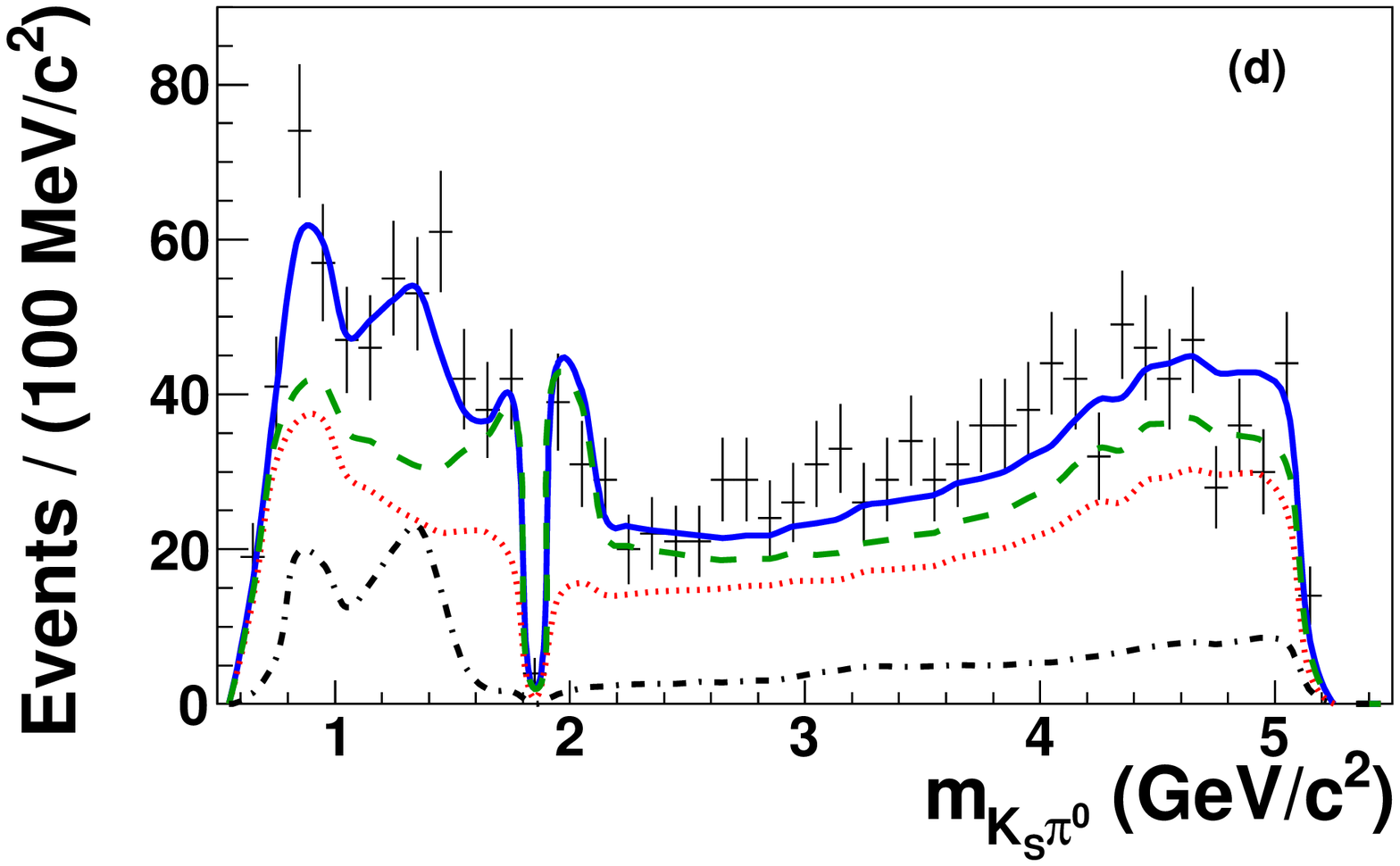}
 \includegraphics[width=.45\textwidth]{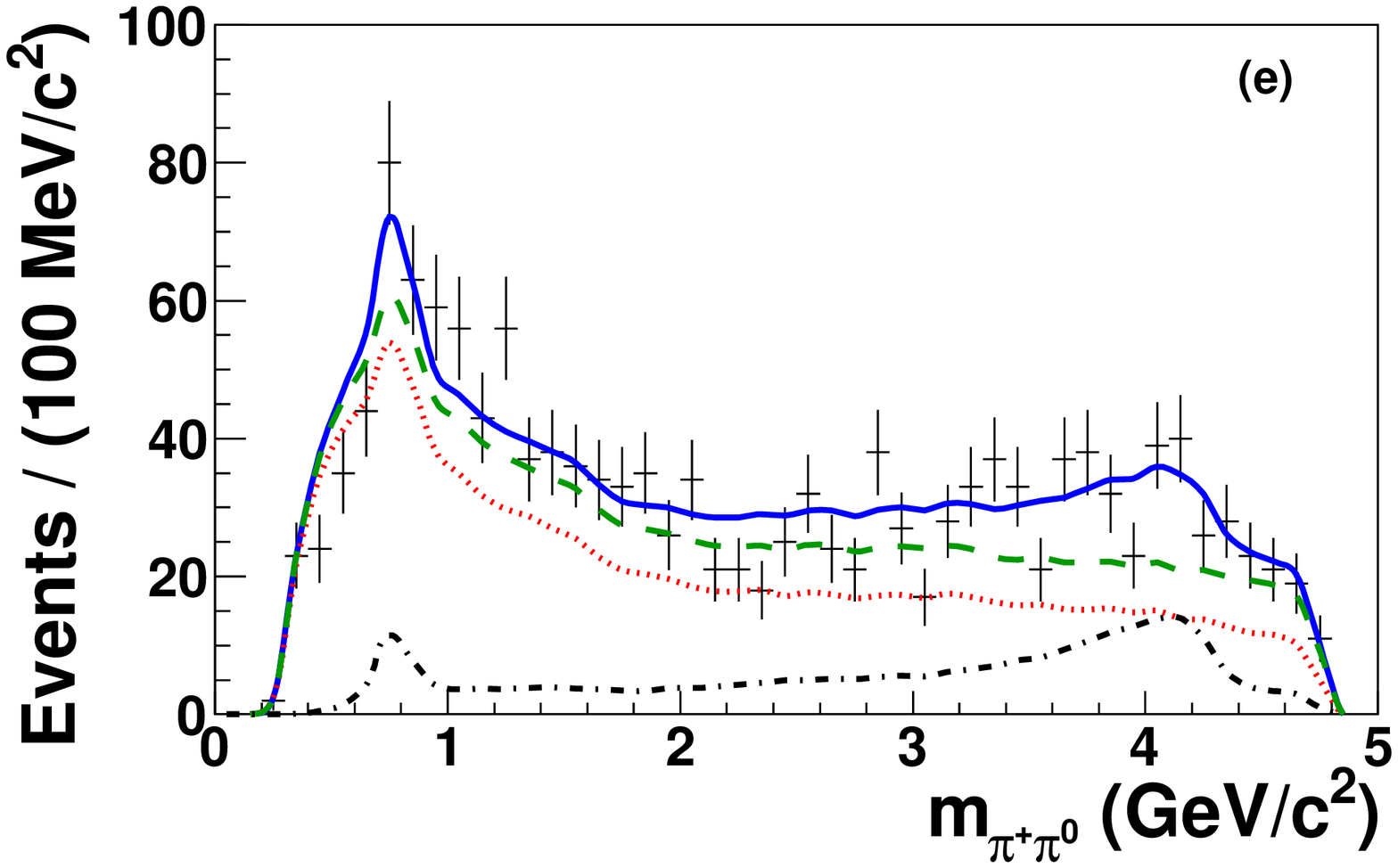}
 \includegraphics[width=.45\textwidth]{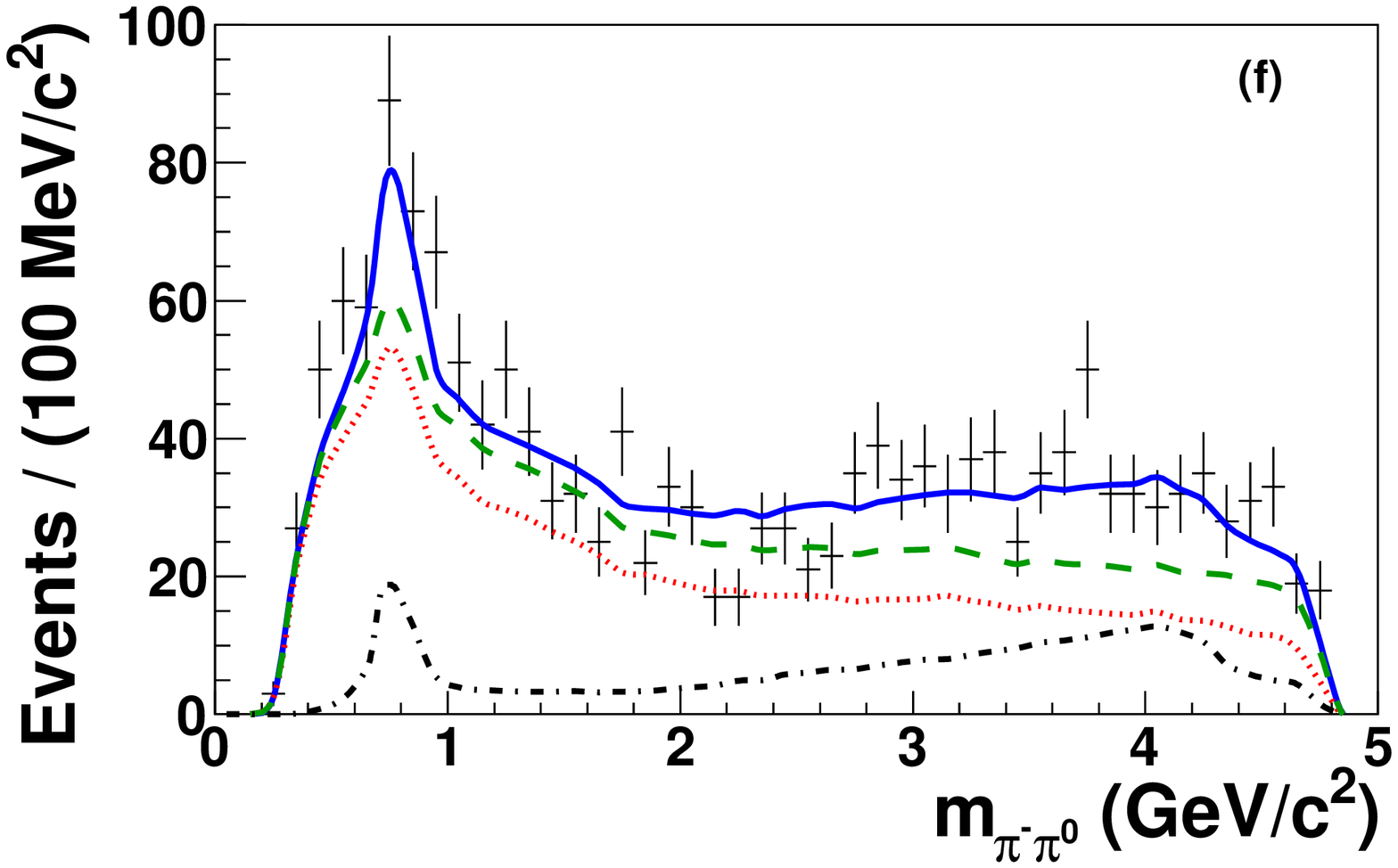}
 \caption{
	 {The \CP\ fit}: Measured distributions and fit projections for
	 $\Bp\to\KS\pip\piz$ (left column) and $\Bm\to\KS\pim\piz$ (right
	 column) candidates;
   (a) \mkspip, 
   (b) \mkspim, 
   (c) \mkspiz (from $\Bp\to\KS\pip\piz$), 
   (d) \mkspiz (from $\Bm\to\KS\pim\piz$), 
   (e) \mpippiz, and
   (f) \mpimpiz.
   The points with error bars correspond to data, the solid (blue) curves
   to the total fit result, the dashed (green) curves to the total
   background  contribution, and the dotted (red) curves to the continuum
   background component. The dash-dotted curves represent the signal
   contribution.  The projected distributions are obtained from
   statistically precise pseudo experiments generated using the fit
   results.  For all distributions in each panel, the signal-to-background
   ratio is increased by applying the tighter selection requirements on
   \mes, \DeltaE, and/or \NN, listed in \tabref{signal-enhance-selections}.  
   }
 \label{fig:signal-project-cpFit}
 \end{figure*}

To determine the overall \CP\ asymmetry as well as the \CP\ asymmetries for the
contributing isobar components, we simultaneously fit the separate \Bp\ and
\Bm\ data samples. The overall \Acp\ value is calculated from the integrals of
the positive and negative signal Dalitz plot distributions.
The $\Delta x$ and $\Delta y$ parameters from Eq.~(\ref{eq:complex-coeff}) are
allowed to vary in the fit for all components except the reference isobar,
for which the $\Delta y$ parameter is fixed to zero (the relative phase of the
\Bp\ and \Bm\ Dalitz plots cannot be determined since they do not
interfere).
To account for possible differences in the reconstruction and particle
identification efficiencies for \Bp\ and \Bm, the efficiency map as a function of the Dalitz plot
position is determined separately for \Bp\ and \Bm.
The asymmetry for the continuum background is allowed to vary in the fit.
The \CP\ asymmetries of the \BB\ backgrounds are expected to be small and so are
fixed to zero in the nominal fit.  They are varied within reasonable
ranges based on world average experimental results~\cite{Beringer:1900zz} in order to
determine the associated systematic uncertainty.

We find an overall \CP\ asymmetry of
$\Acp\left(\BptoKzpippiz\right) = \kpipiAcp$, where the first uncertainty
is statistical, the second is systematic, and the third is due to the signal
model. This is consistent with zero \CP\ asymmetry.  Invariant mass
projections for the fit to data allowing for direct \CP\ violation are shown in
Fig.~\ref{fig:signal-project-cpFit}. 

\renewcommand{\arraystretch}{1.5}
\begin{table}
	\caption{
		Measured branching fractions ${\cal B}$ from a fit to the
		combined \Bpm\ data sample, and \CP\ asymmetries \Acp\
		(Eq.~(\ref{Acp-def})).  The first uncertainty is statistical, the
		second is systematic, and the third is due to the signal
		model.
	}
	\label{tab:results-bf-Acp}
	\begin{tabular}{l|cc}
		\hline
		Decay channel	& ${\cal B}\left(10^{-6}\right)$	& \Acp 			\\ 		
		\hline
		$\Kz\pip\piz$\phantom{\large I}	& \kpipiBFal				& \kpipiAcp		\\
		$\KstarI\pip$	& \kstarIpipBF		 		& \kstarIpipAcp		\\
		$\KstarIp\piz$	& \kstarIppizBF				& \kstarIppizAcp 	\\		
		$\KstarII\pip$  & \kstarIIpipBF				& \kstarIIpipAcp 	\\
		$\KstarIIp\piz$ & \kstarIIppizBF			& \kstarIIppizAcp 	\\
		    $\rhoIp\Kz$	& \rhoIpKzBF				& \rhoIpKzAcp 		\\
		\hline
	\end{tabular}
\end{table}
\renewcommand{\arraystretch}{1.0}

\begin{figure}
	\includegraphics[scale=0.45]{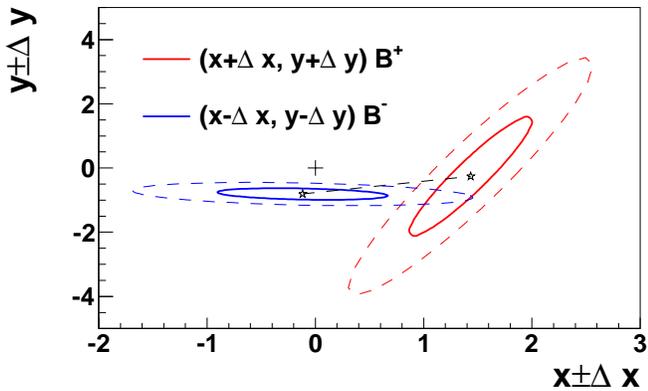}
	\caption{
		\CP\ parameters $(x\pm\Delta x, y\pm\Delta y)$ obtained
		from the fit to data for $\Bpm\to K^{*\pm}(892)\piz$
		resonant decay including the $1$ and $2$ standard deviation contours
		(solid and dashed curves). The contours are estimated by
		calculating the uncertainty and correlation between the two
		\CP\ parameters. The stars indicate the central values of
		the \CP\ parameters and the cross sign the origin of the
		plot.
	}
	\label{fig:cp-ellipse-KstarIppiz}
\end{figure}

\tabref{results-bf-Acp} shows the results for the branching fractions and
\CP\ asymmetries obtained from the fit to data. The first uncertainty is
statistical, the second is systematic, and the third is the uncertainty
associated with the signal model. We observe a
significant asymmetry between the \mkspip\ and \mkspim\ distributions in
the region of the $\KstarIp$ resonance; see
Figs.~\ref{fig:signal-project-cpFit}(a) and (b).  We determine the
statistical significance, $S$, of a non-zero \CP\ asymmetry in
\BptoKstarIppiz\ from the difference between the best-fit value of the
likelihood, ${\cal L}_{\Acp}$, and the value when the \CP\ asymmetry is
fixed to zero, ${\cal L}_{0}$:
\begin{equation}
	S=\sqrt{-2\ln\left({\cal L}_{0}/{\cal L}_{\Acp}\right)}.
\end{equation}
Using this method, we measure a statistical significance of $3.6$ standard
deviations for a non-zero \Acp\ in \BptoKstarIppiz.  We obtain a consistent
result of $3.7$ standard deviations for the statistical significance by
dividing the central value of the \CP\ asymmetry by the statistical
uncertainty, indicating that the log-likelihood function is close to parabolic.
Figure~\ref{fig:cp-ellipse-KstarIppiz} displays the contours in the complex plane
of the coefficients $c=\left(x+\Delta x,y+\Delta y\right)$, defined in
Eq.~(\ref{isobar-formalism-A}), for \BptoKstarIppiz\ decays, and of
$\bar{c}=\left(x-\Delta x,y-\Delta y\right)$, defined in
Eq.~(\ref{isobar-formalism-Abar}), for \BmtoKstarImpiz\ decays. For other
resonances the \CP\ asymmetry is within $2$ standard deviations of zero.

\begin{table*}
	\caption{
		 Results for the relative phases $\phi$ obtained from the
		 combined \Bpm\ fit, the \CP\ amplitudes
		 $A_{+}$ and $A_{-}$, and the \CP\ phases $\phi_{+}$ and
		 $\phi_{-}$ obtained from the \CP\ fit.
		 All parameters are measured relative to the
		 $\Bpm\to\KstarI\pipm$ reference amplitude.  The first
		 uncertainty is
		 statistical, the second is systematic, and the third is
		 due to the signal model. Note that for the \CP\ phases of
		 all contributions except for $\Bpm\to\KpiSwave\pipm$, only
		 statistical uncertainties are quoted. 
	}
	\label{tab:results-ph-amp}
	\begin{tabular}{l|cccccc}
		\hline
		Isobar		 & $\phi$ $(^{\circ})$	&  $A_{+}$	& $A_{-}$ 		& $\phi_{+}$ $(^{\circ})$ & $\phi_{-}$ $(^{\circ})$\\
		\hline
		$\KstarIpm\piz$	 &  \kstarIppizPhNonCP	& \kstarIppizA	& \kstarIppizAbar	& \kstarIppizPh & \kstarIppizPhBar 	\\		
	       $\KpiSwave\pipm$    & \kstarIIpipPhNonCP	& \kstarIIpipA	& \kstarIIpipAbar	& \kstarIIpipPh & \kstarIIpipPhBar\\
		$\KpiSwavepm\piz$  &  \kstarIIppizPhNonCP	& \kstarIIppizA	& \kstarIIppizAbar	& \kstarIIppizPh & \kstarIIppizPhBar 	\\
		    $\rhoIpm\Kz$	 &  \rhoIpKzPhNonCP	& \rhoIpKzA 	& \rhoIpKzAbar		& \rhopKzPh 	& \rhopKzPhBar 		\\
		\hline
	\end{tabular}
\end{table*}

We also express the complex isobar coefficients $c$ and $\bar{c}$ of
Eq.~(\ref{eq:complex-coeff}) in terms of amplitudes and phases,
\begin{eqnarray}
	c&=&A_{+}e^{i\phi_{+}}\,, \\
  \bar{c}&=&A_{-}e^{-i\phi_{-}}\,.
\end{eqnarray}
\tabref{results-ph-amp} presents the results, measured with respect to the
$\Bpm\to\KstarI\pipm$ reference amplitude. The statistical uncertainties of
the separate \Bp\ and \Bm\ decay amplitudes, $A_{+}$ and $A_{-}$, vary
between $0.1$ and $0.3$. We thus obtain significant statistical precision
for these terms. With respect to the phases, $\phi_{+}$ and $\phi_{-}$,
only the \KpiSwave\ amplitude yields a statistically precise result. For
the other amplitudes, the statistical uncertainty ranges between
$70^{\circ}$ and $170^{\circ}$, and only the statistical uncertainty is
quoted. For the more precisely determined variables, systematic
uncertainties are evaluated as well. For the phases of the
$\Bpm\to\KpiSwavepm\piz$ decays relative to the $\Bpm\to\KstarIpm\piz$
amplitude, we obtain
\small
\begin{eqnarray}
	\phi_{+}\left(\KpiSwavep\piz\right) - \phi_{+}\left(\KstarIp\piz\right) = \left(\deltaKstarppizPh\right)^{\circ}, \nonumber \\
	\\
	 \phi_{-}\left(\KpiSwavem\piz\right) -
	 \phi_{-}\left(\KstarIm\piz\right)=
	 \left(\deltaKstarppizPhBar\right)^{\circ}.\nonumber
\end{eqnarray}
\normalsize

\section{Systematic uncertainties}
\label{sec:systematics}

We evaluate systematic uncertainties to account for effects that could
affect the branching fractions, phases, and asymmetries, by varying the
fixed parameters.  The systematic uncertainties described in this section
are summarized in Tables~\ref{tab:systematics-bf} through
\ref{tab:fit-cpPh-sys} of Appendix~\ref{sec:sys-tables}. 

The uncertainties associated with the branching fractions are listed in
\tabref{systematics-bf}.  To estimate the uncertainty related to the 
modeling of the SCF PDFs,  we implement a simpler model
consisting of only four regions in the Dalitz plot. The PDFs are redefined
using MC events to match the distributions found in the newly defined regions. We
then fit the data using the new SCF model and take the uncertainties to be
the change in the fit parameters compared to those obtained from the
nominal fit to data. All relative systematic uncertainties due to the SCF
\mes\ and \DeltaE\ PDFs range from approximately $1\%$ to $4\%$, except for the
relative systematic uncertainty for the \BptorhopKz\ decay, which is
$7.5\%$. This is consistent with expectations from simulation that more
than half the \BptorhopKz\ events are due to SCF. 

The uncertainties associated with the number of \BB\ background events are evaluated by
varying the estimates within their uncertainties, which are primarily due
to uncertainties in the branching fractions.  The uncertainties related to the \BB\
background \mes, \DeltaE, and \NN\ PDFs are accounted for by varying the
histogram bin contents according to their statistical uncertainties.
The uncertainty is then taken as the RMS of the distribution of the
difference in the fit parameters. The uncertainties related to the limited
statistical precision of the MC and data-sideband samples are similarly
accounted for by varying the results in the corresponding histogram bins by
their uncertainties. 

The uncertainty in the \NN\ histogram PDFs for correctly reconstructed and
SCF signal events is determined by varying the bin contents in
accordance with the observed data/MC difference.  For correctly
reconstructed signal events, the tails of the asymmetric Gaussian PDFs for \mes\
and \DeltaE\ are fixed. To account
for an associated uncertainty, we allow the relevant parameters to vary in a fit to data
and use the variation in the fit parameters to define the uncertainty.

To validate the fitting procedure, 500 MC pseudo experiments are
generated, using the PDFs with parameter values found from the fit to data.
Small fit biases are found for some of the fit parameters and are included
in the systematic uncertainties.

We also account for uncertainties in the following parameters describing 
the signal model: the mass and width of each resonance and the value of the
Blatt-Weisskopf barrier radius.  The associated uncertainties are
determined by varying the parameters within their uncertainties (some of
which are given in \tabref{signal-model-description}) and refitting. 

The uncertainties in the branching fractions related to particle
identification, tracking efficiency, and the total number of \BB\ events
are 1.0\%, 1.0\%, and 0.6\%, respectively. We estimate systematic
uncertainties in the branching fractions associated with the \piz\ and \KS\
reconstruction efficiencies to be 1.0\% and 1.1\%, respectively. 

Uncertainties from all the above sources are added in quadrature to yield
the total systematic uncertainties, which are listed in \tabref{systematics-bf}. 

We determine changes in the branching fractions, $\Delta{\cal B}$, when 
the signal model is varied. The systematic uncertainties in
the branching fractions due to the $(K\pi)^{*0/+}_{0}$ parametrization are
estimated by replacing the LASS model with another phenomenologically
inspired parametrization~\cite{ElBennich:2009da}. We take the differences
in branching fractions with respect to the nominal fit as the systematic
uncertainty. This is the largest contribution to the uncertainty due to the
model.  Another uncertainty reflects any changes in the fit parameters for
the nominal model when including components that are omitted in the nominal
fit, such as the \rhoIIp, \KstarIII, and \KstarIIIp. Positive and negative
variations are added separately in quadrature to obtain the systematic
uncertainties due to the signal model, listed in \tabref{systematics-bf}.

We determine systematic uncertainties in the phases averaged over \Bp\ and
\Bm\ decays from the same sources as considered for the branching
fractions. The variations in the phases are measured relative to the
$\KstarI\pip$ amplitude. Since the differences between positive and
negative shifts in the phases, shown in \tabref{fit-ph-sys}, are large in
some cases, we quote for those phase shifts asymmetric systematic
uncertainties.

Reconstruction and particle identification efficiencies cancel to first
order in the fit to \CP\ asymmetries; therefore the only uncertainties that are
included for \Acp\ are those coming from the fit and signal model.  In addition to
this, we do not evaluate any of the uncertainties that are found to be
negligible for the branching fractions.

An additional uncertainty for \Acp\ arises from having fixed the \CP\
asymmetries for individual \BB\ background components to the mean asymmetry
averaged over all such components.  We take the largest variation of each
background asymmetry as the corresponding uncertainty. 

The uncertainty related to the efficiency model is determined by exchanging the
efficiency maps for the positive and negative Dalitz plots and refitting the data.
We then take the difference in \CP\ asymmetry with respect to the nominal fit as the
uncertainty.

We list in \tabref{fit-acp-sys} the systematic uncertainties associated with
the signal \CP\ asymmetries and the variations in the asymmetry due to
changes in the signal composition. 

We evaluate systematic uncertainties for the \CP\ amplitudes and \CP\ phases
from the same sources as for the \CP\ asymmetries. We list
the variations to the amplitudes $A_{+}$ in \tabref{fit-amp-sys-A} and to
the amplitudes
$A_{-}$ in \tabref{fit-amp-sys-Abar}, including the uncertainties due to
changes to the signal model.  \tabref{fit-cpPh-sys} lists the systematic
variations and model uncertainties for
$\phi_{+(-)}\left(\left(K\pi\right)^{*0}_{0}\pi^{+(-)}-K^{*}(892)^{0}\pi^{+(-)}\right)$
and
$\phi_{+(-)}\left(\left(K\pi\right)^{*+(-)}_{0}\pi^{0}-K^{*}(892)^{+(-)}\pi^{0}\right)$.

\section{Summary and Conclusion}
\label{sec:conclusion}

\begin{table*}
	\caption{
		Combined measurements of branching fractions and \CP\
		asymmetries from \BptoKspippiz\ (this analysis) and
		from \babar\ analyses of \BptoKstarIpip\ and
		\BptoKstarIIpip\ from \BptoKppimpip~\cite{Aubert:2008bj},
		and of \BptoKstarIppiz\ from
		\BptoKppizpiz~\cite{BABAR:2011aaa}.  The first uncertainty
		is statistical and the second is systematic.
	}
	\label{tab:results-bf-Acp-combined}
	\begin{tabular}{l|cc}
		\hline
		Decay channel	 & ${\cal B}\left(10^{-6}\right)$ 	& \Acp 	 	\\
		\hline
		$\KstarI\pip$\phantom{\large I}	 & \kstarIpipBFComb	& \kstarIpipAcpComb	\\
		$\KstarIp\piz$	 & \kstarIppizBFComb			& \kstarIppizAcpComb 	\\		
		$\KstarII\pip$   & \kstarIIpipBFComb			& \kstarIIpipAcpComb 	\\
		\hline
	\end{tabular}
\end{table*}

The measured branching fractions and \CP\ asymmetries are summarized in
\tabref{results-bf-Acp}, and the amplitude and phase values in
\tabref{results-ph-amp}, including statistical, systematic, and model
uncertainties. We have measured for the first time the branching
fraction and \CP\ asymmetry for the decay \BptoKzpippiz. We obtain first
evidence for direct \CP\ violation in the intermediate decay
\BptoKstarIppiz, with a total significance of $3.4$ standard deviations determined by adding
statistical, systematic, and signal-model uncertainties in quadrature and dividing the
measured \Acp\ by the total uncertainty.

In addition, we have measured the branching fractions, \CP\ asymmetries, and
relative \CP-averaged phase  values of the decays \BptoKstarIpip,
\BptoKstarIppiz, \BptoKstarIIpip, \BptoKstarIIppiz, and \BptorhopKz.
The results for the branching fractions and \CP\ asymmetries for
\BptoKstarIpip\ are consistent with the previous
measurement from \BptoKppimpip\ decays by the Belle and \babar\
Collaborations and the results for \BptoKstarIIpip\ are within two standard deviations from
the previous \babar\ measurement~\cite{Aubert:2008bj,Garmash:2005rv}. The branching fraction
for \BptoKstarIppiz\ are consistent with the previous measurements from the
\babar\ Collaboration for the decay mode $\Bp\to\Kp\piz\piz$ and the result
for \Acp\ is within $2$ standard deviations of the previous
measurement~\cite{BABAR:2011aaa}.  The branching fraction and \Acp\ results for
\BptorhopKz\ supersede the previous \babar\ measurements~\cite{Aubert:2007mb}.  The \CP\
asymmetries of \BptoKstarIpip, \BptoKstarIIpip, and \BptorhopKz\ are all
consistent with zero, as expected. We obtain the first measurements of the
branching fraction and \CP\ asymmetry for \BptoKstarIIppiz, with a
significance of $5.4$ standard deviations for the branching fraction. 

We combine our results for the branching fractions and \CP\ asymmetries of
\BptoKstarIpip, \BptoKstarIIpip, and \BptoKstarIppiz\ with the previous
\babar\ measurements. The statistical uncertainties and all systematic
uncertainties for the \CP\ asymmetries are uncorrelated between the
measurements. For the branching fractions, we account for
possible correlations when combining the systematic uncertainties. 
If the systematic uncertainties are asymmetric, the average systematic
uncertainty is calculated from the largest limit. The combined results from
\babar\ for these decay modes are presented in
\tabref{results-bf-Acp-combined}.

Using the world average value for direct \CP\ violation in
$\Bz\to\KstarIp\pim$~\cite{Amhis:2012bh} and the final \babar\ result for
direct \CP\ violation in \BptoKstarIppiz, we calculate $\Delta\Acp$ for the
$K^{*}\pi$ system to be
\begin{eqnarray}
	\Delta\Acp\left(K^{*}\pi\right)&=&\Acp\left(K^{*+}\piz\right)-\Acp\left(K^{*+}\pim\right) \nonumber\\
					&=&-0.16\pm0.13.	
\end{eqnarray}

Thus the value of $\Delta\Acp$ in $\Kstar\pi$ is found to be consistent with zero.  
The uncertainty in the $\Delta\Acp(\Kstar\pi)$ result remains large,
rendering the comparison to $\Delta\Acp(K\pi)$, given in
Eq.~(\ref{eq:deltaAcp-Kpi}), inconclusive at present and motivating
improved determinations in future experiments.

\renewcommand{\arraystretch}{1}

\section{Acknowledgements}

We are grateful for the 
extraordinary contributions of our \pep2\ colleagues in
achieving the excellent luminosity and machine conditions
that have made this work possible.
The success of this project also relies critically on the 
expertise and dedication of the computing organizations that 
support \babar.
The collaborating institutions wish to thank 
SLAC for its support and the kind hospitality extended to them. 
This work is supported by the
US Department of Energy
and National Science Foundation, the
Natural Sciences and Engineering Research Council (Canada),
the Commissariat \`a l'Energie Atomique and
Institut National de Physique Nucl\'eaire et de Physique des Particules
(France), the
Bundesministerium f\"ur Bildung und Forschung and
Deutsche Forschungsgemeinschaft
(Germany), the
Istituto Nazionale di Fisica Nucleare (Italy),
the Foundation for Fundamental Research on Matter (The Netherlands),
the Research Council of Norway, the
Ministry of Education and Science of the Russian Federation, 
Ministerio de Econom\'{\i}a y Competitividad (Spain), the
Science and Technology Facilities Council (United Kingdom),
and the Binational Science Foundation (U.S.-Israel).
Individuals have received support from 
the Marie-Curie IEF program (European Union) and the A. P. Sloan Foundation (USA). 


\bibliographystyle{apsrev}
\bibliography{references}

\newpage

\appendix

\section{Tables of Systematic and Model Uncertainties}
\label{sec:sys-tables}

\tabref{systematics-bf} lists the uncertainties in the branching fractions due to systematic effects,
efficiency corrections, and changes to the signal model. \tabref{fit-ph-sys}
lists uncertainties in the relative phase values (for \Bp\ and \Bm\ decays
combined) due to systematic effects and changes to the signal
model. Tables \ref{tab:fit-acp-sys}, \ref{tab:fit-amp-sys-A},
\ref{tab:fit-amp-sys-Abar}, and \ref{tab:fit-cpPh-sys} list the systematic
and signal model uncertainties for the \CP\ asymmetries, the amplitudes for
the \Bp\ and \Bm\ Dalitz plots, $A_{+}$ and $A_{-}$, respectively, and the
corresponding phases $\phi_{+}$ and $\phi_{-}$ for the \BptoKstarIIpip\
amplitude relative to that for \BptoKstarIpip, and the phase values
for the \BptoKpiSwaveppiz\ amplitude relative to that for \BptoKstarIppiz.

\begin{table*}
  \caption{
    {Combined \Bpm\ fit}: Systematic uncertainties for the branching fraction
    measurements, including uncertainties due to the signal model.
  }
  \label{tab:systematics-bf}
  \begin{tabular}{l|cccccc}
    \hline
    		& \multicolumn{6}{c}{Relative Variations of branching fraction (\%)} \\
    \backslashbox{Source}{Resonant contribution}   & Inclusive & \KstarI\ & \KstarIp\	& \KstarII & \KstarIIp		& \rhoIp	\\
    \hline
    Correctly reconstructed \mes\ and \DeltaE\ PDF (fixed parameters)	& $0.8$	& $1.1$	& $0.6$	& $1.1$	& $0.7$ & $1.2$	\\
    Correctly reconstructed and self crossfeed signal \NN\ PDFs &  $3.3$	& $3.3$	& $3.4$	& $3.4$	& $4.2$ & $4.0$		\\
    Self crossfeed signal \mes\  and \DeltaE\ PDF models &  $3.0$	& $4.3$	& $3.1$	& $1.3$	& $1.8$ & $7.5$		\\
    Fit bias 				      &  $0.3$	& $0.9$	& $0.6$	& $0.5$	& $0.7$ & $0.9$			\\
    \BB\ background \mes, \DeltaE\ and \NN\ PDFs &  $0.3$	& $0.4$	& $0.2$	& $0.3$	& $0.5$ & $0.6$	\\
    \BB\ background yields & $0.7$	& $1.2$	& $0.6$	& $0.9$	& $2.0$ & $1.8$	\\
    Background model in Dalitz plot	&  $1.5$	& $3.7$	& $2.8$	& $2.8$	& $2.7$ & $3.5$	\\
    Signal efficiency model		&  $0.3$	& $1.8$	& $1.0$	& $0.4$	& $0.4$ & $0.8$	\\
		$K^{*}(892)$ mass and width	& $0.1$	& $0.7$	& $0.3$	& $0.1$	& $0.2$ & $0.1$		\\
		$K^{*}_{0}(1430)$ mass and width & $3.2$	& $3.8$	& $2.1$	& $8.1$	& $5.5$ & $4.0$		\\
		\rhoIp\ mass and width 		& $<0.1$	& $0.2$	& $0.1$	& $0.1$	& $0.2$ & $0.3$	\\
		Blatt-Weisskopf radius			& $2.3$	& $4.4$	& $2.9$	& $7.4$	& $2.9$ & $3.7$	\\
    \hline
		Subtotal			& $6.3$	& $9.1$	& $6.6$	& $12.0$ & $8.5$ & $11.0$	\\
    \hline
    Neutral pion efficiency & $1.0$ & $1.0$ & $1.0$ & $1.0$ & $1.0$ & $1.0$ \\
    \KS\ efficiency & $1.1$ & $1.1$ & $1.1$ & $1.1$ & $1.1$ & $1.1$ \\
    Charged particle identification efficiency & $1.0$ & $1.0$ & $1.0$ & $1.0$ & $1.0$ & $1.0$ \\
    Tracking efficiency & $1.0$ & $1.0$ & $1.0$ & $1.0$ & $1.0$ & $1.0$ \\
    \nbb & $0.6$ & $0.6$ & $0.6$ & $0.6$ & $0.6$ & $0.6$ \\
    \hline
    Total 		& $6.6$ & $9.4$	& $7.0$	& $12.2$	& $8.7$ & $11.2$	\\
    \hline
    \hline
    Changes due to signal model	& \multicolumn{6}{c}{$\Delta{\cal B}\left(10^{-6}\right)$} \\
    \hline
    \KpiSwave/\KpiSwavep\ parametrization		& $+8.0$ & $-0.3$		& $-0.3$		& --	& -- 	& $-1.4$	\\
					      \rhoIIp\			& $+2.3$	& $+0.3$ & $-0.4$		& $+2.7$		& $-0.8$ & $-2.0$	\\
	\KstarIII\ and \KstarIIIp\		& $+1.4$	& $-0.3$		& $+0.3$	& $-2.6$			& $-0.8$ & $-0.3$	\\
			 \KstarIV\ and \KstarIVp\	& $+1.8$	& $-0.1$		& $-0.1$		& $+0.6$ & $-1.4$ & $-0.2$	\\
    \hline
    Total $(+)$				& $+8.6$	& $+0.3$		&	$+0.3$	& $+2.7$ & $+0.0$ & $+0.0$	\\
    Total $(-)$				& $-0.0$	& $-0.4$		&	$-0.5$	& $-2.6$ & $-1.8$ & $-2.4$	\\
    \hline
    \end{tabular}
    \end{table*}

\renewcommand{\arraystretch}{1.5}
\begin{table*}
	\caption{{Combined \Bpm\ fit}: Systematic uncertainties due to the fit model,
		fixed shapes in the parametrization, and changes to the
		signal model for the relative phases (in degrees) measured relative
		to the \KstarI\ amplitude.
	}
	\label{tab:fit-ph-sys}
	\begin{tabular}{l|cccc}
		\hline
			& \multicolumn{4}{c}{Systematic Variations ($^{\circ}$)}	\\
		\backslashbox{Systematic}{Resonant contribution} &  $\KstarIp\piz$ 	& $\KstarII\pip$	& $\KstarIIp\piz$	& $\rhoIp\KS$	\\
		\hline
		Self crossfeed PDFs and mapping	& $7.4$	& $1.4$	& $9.6$	& $6.7$		\\
		Correctly reconstructed and self crossfeed \NN\ PDFs    & $1.4$	& $0.9$	& $1.3$	& $1.3$	\\
		\BB\ background yields 		& $1.7$	& $0.5$	& $1.8$	& $2.1$	\\
		Correctly reconstructed \mes\ and \DeltaE\ PDF	& $0.4$	& $0.2$	& $0.9$	& $1.2$ \\
		Background DP PDF	& $5.7$	& $2.2$	& $4.5$	& $5.4$	\\
		\BB\ background \mes, \DeltaE, \NN\ PDFs	& $0.3$	& $0.2$	& $0.3$	& $0.3$\\
		Signal efficiency model& $0.5$	& $0.4$	& $0.7$	& $0.7$	\\
		Fit bias		& $9.3$	& $7.8$	& $7.3$	& $23.5$\\
		$K^{*}(892)$ mass 	& $1.4$	& $0.2$	& $1.2$	& $1.4$	\\
		$K^{*}(892)$ width	& $0.3$	& $0.1$	& $0.1$	& $<0.1$	\\
    $K^{*}_{0}(1430)$ mass  & $^{+43}_{-33}$	& $^{+5.2}_{-4.7}$	& $^{+48}_{-37}$	& $^{+46}_{-36}$\\
		$K^{*}_{0}(1430)$ width & $5.2$	& $4.0$	& $5.7$	& $14.6$\\
		\rhoIp\ mass 		& $0.3$	& $0.1$	& $0.4$	& $1.0$	\\
		\rhoIp\ width 		& $0.6$	& $0.1$	& $0.6$	& $0.2$	\\
		Blatt-Weisskopf radius 	& $^{+15}_{-2}$	& $^{+1.0}_{-1.4}$	& $^{+17}_{-3}$	& $^{+0.3}_{-0.5}$	\\
		\hline
		Total			& $^{+48}_{-36}$	& $^{+11}_{-11}$	& $^{+53}_{-40}$	& $^{+55}_{-47}$ \\
		\hline
		\hline
    \multicolumn{5}{c}{Changes due to signal model} \\
		\hline
		\KpiSwave/\KpiSwavep\ parametrization		& $-67.0$ &  --		& -- 	 & $-60.3$	\\
		\rhoIIp\			& $-18.4$	& $-2.8$ & $-11.8$		& $-27.4$\\
		 $K^{*}_{2}(1430)$		& $+7.8$	& $-3.1$ & $+5.5$		& $+11.3$ \\
			      $K^{*}(1680)$	& $-7.8$	& $-4.9$ & $-12.9$		& $+11.8$ \\
		\hline
		Total $(+)$			& $+7.8$	& $+0.0$	& $+5.5$	& $+16.4$	\\
		Total $(-)$			& $-69.9$	& $-6.5$ & $-17.5$	& $-66.2$	\\
		\hline
	\end{tabular}
\end{table*}
\renewcommand{\arraystretch}{1.0}

\begin{table*}
	\caption{Contributions to the uncertainties in the \CP\ asymmetries for the
		overall and resonant isobar contributions, including
		uncertainties due to changes to the signal model.
	}
	\label{tab:fit-acp-sys}
	\begin{tabular}{l|cccccc}
		\hline
			& \multicolumn{6}{c}{Systematic Variations of \Acp\ (\%) }	\\
		\backslashbox{Systematic}{Resonant contribution} & Inclusive & $\KstarI\pip$	& $\KstarIp\piz$ 	& $\KstarII\pip$	& $\KstarIIp\piz$	& $\rhoIp\KS$	\\
		\hline
		Self crossfeed PDFs and mapping	& $2.0$	& $6.0$	& $1.0$	& $1.2$	& $0.9$ & $5.0$		\\
		Correctly reconstructed and self crossfeed \NN\ PDFs    & $0.7$	& $1.7$	& $1.8$	& $1.6$	& $1.1$ & $2.8$		\\
		\BB\ background asymmetries	& $2.5$	& $1.4$	& $1.7$	& $1.8$	& $7.5$ & $2.3$	\\
		Background DP PDF	& $0.7$	& $2.7$	& $2.1$	& $2.8$	& $2.0$ & $2.5$	\\
		\BB\ background \mes, \DeltaE, \NN\ PDFs	& $0.2$	& $0.1$	& $0.2$	& $0.3$	& $0.5$ & $0.4$	\\
		Signal efficiency model	& $0.2$	& $3.9$	& $2.1$	& $0.1$	& $0.9$ & $1.3$	\\
		Fit bias		& $0.3$	& $1.4$	& $0.8$	& $1.9$	& $1.0$ & $1.2$	\\
		$K^{*}(892)$ mass and width	& $0.1$	& $0.2$	& $0.4$	& $0.2$	& $0.2$ & $0.2$		\\
		$K^{*}_{0}(1430)$ mass and width & $1.1$	& $4.4$	& $0.5$	& $3.0$	& $2.8$ & $2.2$		\\
		\rhoIp\ mass and width 		& $<0.1$	& $0.1$	& $0.1$	& $0.2$	& $0.1$ & $0.1$	\\
		Blatt-Weisskopf radius		& $<0.1$	& $0.9$	& $0.3$	& $0.8$	& $1.1$ & $0.5$	\\
		\hline
		Total			& $3.4$	& $8.1$	& $4.1$	& $4.3$ &	$8.0$ & $6.9$	\\
		\hline
		\hline
    \multicolumn{7}{c}{Changes due to signal model} \\
		\hline
		\KpiSwave/\KpiSwavep\ parametrization		& $-0.7$
						       &	$+6.2$	&
						 $+2.0$	& --	&	-- 	 &	$-8.1$	\\
		\rhoIIp\			& $+3.3$		&
		     $+1.5$	& 	$-3.4$		& $-10.5$		&$-11.6$ 	& $-21.3$	\\
		 $K^{*}_{2}(1430)$		& $-0.2$		&
			     $+5.7$		& $-1.5$		&
			     $-7.5$		& $-2.7$ 	& $+14.4$	\\
		$K^{*}(1680)$			& $-2.2$		&
			 $+6.3$		& $+0.5$		& $+4.8$
				 & $-1.9$ 	& $+12.3$	\\
		\hline
		Total $(+)$		& $+2.4$	& $+0.0$	& $+3.7$	& $+13.0$	& $+12.0$ & $+22.8$	\\
		Total $(-)$		& $-3.3$	& $-10.6$	& $-2.0$	& $-4.8$	& $-0.0$ & $-19.0$	\\
		\hline
	\end{tabular}
\end{table*}

\begin{table*}
	\caption{Variations in the \CP\ amplitude, $A_{+}$, including
		uncertainties due to changes to the signal model. In the
		fits, the amplitudes are measured relative to the \KstarI\
		amplitude.
	}
	\label{tab:fit-amp-sys-A}
	\begin{tabular}{l|cccc}
		\hline
		& \multicolumn{4}{c|}{Variation of $A_{+}$ } 	\\
		\backslashbox{Systematic}{Resonant contribution} & $\KstarIp\piz$ 	& $\KpiSwave\pip$	& $\KpiSwavep\piz$ & $\rhoIp\KS$ \\
		\hline
		Self crossfeed PDFs and mapping	& $0.02$	& $0.02$	& $0.04$	& $0.02$	\\
      		Correctly reconstructed and self crossfeed \NN\ PDFs    & $0.01$	& $0.03$	& $0.02$	& $0.02$ 	\\
		\BB\ background asymmetries & $0.01$	& $0.03$	& $0.07$	& $0.02$	\\
		Background DP PDF	& $0.02$	& $0.06$	& $0.04$	& $0.04$	\\
		\BB\ background \mes, \DeltaE, \NN\ PDFs	& $<0.01$	& $0.01$	& $0.01$	& $<0.01$	\\
		Signal efficiency model	& $0.01$	& $0.01$	& $<0.01$	& $0.01$	\\
		Fit bias		& $0.01$	& $0.03$	& $0.02$	& $0.03$	\\
		$K^{*}(892)$ mass 	& $0.01$	& $0.01$	& $0.01$	& $<0.01$	\\
		$K^{*}(892)$ width	& $<0.01$	& $0.01$	& $0.01$	& $<0.1$	\\
		$(K\pi)^{*}_{0}$ mass  	& $0.02$	& $0.02$	& $0.09$ 	& $0.06$	\\
		$(K\pi)^{*}_{0}$ width 	& $0.02$	& $0.06$	& $0.01$ 	& $0.02$ 		\\
		\rhoIp\ mass 		& $<0.01$	& $<0.01$	& $<0.01$	& $<0.01$ 		\\
		\rhoIp\ width 		& $<0.01$	& $<0.01$	& $<0.01$	& $<0.01$		\\
		Blatt-Weisskopf radius 	& $0.02$	& $0.03$	& $0.02$ 	& $0.02$	\\
		\hline
		Total			& $0.05$	& $0.11$	& $0.13$	& $0.09$ 	\\
		\hline
		\hline
    \multicolumn{5}{c}{Changes due to signal model} \\
		\hline
		\KpiSwave/\KpiSwavep\ parametrization		& $-0.03$ & $--$		& $--$	& $-0.17$  \\
      		\rhoIIp\					& $<0.01$	& $-0.01$ 		& $-0.02$	& $-0.12$ 		\\
		$K^{*}_{2}(1430)$				& $0.05$	& $-0.08$ 		& $-0.08$	& $-0.04$	 \\
		$K^{*}(1680)$					& $-0.02$	& $0.07$ 		& $-0.05$	& $-0.05$		\\
		\hline
		Total $(+)$			& $+0.05$	& $+0.07$ & $+0.00$	& $+0.00$ 		\\
		Total $(-)$			& $-0.03$	& $-0.12$ & $-0.10$	& $-0.21$ 		\\
		\hline
	\end{tabular}
\end{table*}

\begin{table*}
	\caption{Variations in the \CP\ amplitude, $A_{-}$, including
		uncertainties due to changes to the signal model. In the
		fits, the amplitudes are measured relative to the \KstarI\
		amplitude.
	}
	\label{tab:fit-amp-sys-Abar}
	\begin{tabular}{l|cccc}
		\hline
		&	\multicolumn{4}{c}{Variation of $A_{-}$ } 	\\
		\backslashbox{Systematic}{Resonant contribution} &  $\KstarIm\piz$ 	& $\KpiSwave\pim$	& $\KpiSwavem\piz$	& $\rhoIm\KS$	\\
		\hline
		Self crossfeed PDFs and mapping	&  $<0.01$ 	& $0.05$ 	& $0.04$ 	& $0.05$ \\
      		Correctly reconstructed and self crossfeed \NN\ PDFs    &  $0.02$ 	& $0.02$ 	& $0.02$	& $0.03$ 	\\
		\BB\ background asymmetries &  $0.01$ 	& $0.01$ 	&$0.06$	& $0.02$ \\
		Background DP PDF	&  $0.02$ 	& $0.05$ 	& $0.04$	& $0.03$	\\
		\BB\ background \mes, \DeltaE, \NN\ PDFs	&  $<0.01$ 	& $<0.01$ 	& $<0.01$	& $<0.01$	\\
		Signal efficiency model	&  $0.02$ 	& $<0.01$ 	& $0.02$	& $0.02$	\\
		Fit bias		&  $0.02$ 	& $<0.01$	& $0.02$	& $0.02$	\\
		$K^{*}(892)$ mass 	&  $<0.01$ 	& $0.01$	& $0.01$	& $<0.01$ \\
		$K^{*}(892)$ width	&  $<0.01$ 	& $0.01$	& $0.01$	& $0.01$	\\
		$(K\pi)^{*}_{0}$ mass  	&  $0.01$	& $0.08$	& $0.08$	& $0.04$	\\
		$(K\pi)^{*}_{0}$ width 	&  $0.01$	& $0.07$ 	& $0.05$	& $0.03$	\\
		\rhoIp\ mass 		&  $<0.01$	& $<0.01$	& $<0.01$	& $<0.01$		\\
		\rhoIp\ width 		&  $<0.01$	& $<0.01$	& $<0.01$	& $<0.01$	\\
		Blatt-Weisskopf radius 	&  $0.01$	& $0.01$	& $0.05$	& $0.01$	\\
		\hline
		Total			&   $0.05$	& $0.13$	& $0.14$	& $0.09$	\\
		\hline
		\hline
    \multicolumn{5}{c}{Changes due to signal model} \\
		\hline
		\KpiSwave/\KpiSwavep\ parametrization		&  $-0.04$ 	& $--$ 	& $--$ 	& $-0.09$ \\
      		\rhoIIp\					&  $0.04$ 	& $0.12$ 	& $0.22$	& $0.22$ 	\\
		$K^{*}_{2}(1430)$				&  $0.05$ 	& $0.07$ 	&$0.05$		& $0.03$ \\
		$K^{*}(1680)$					&  $-0.01$ 	& $-0.02$ 	& $-0.02$	& $-0.04$	\\
		\hline
		Total $(+)$			&   $+0.06$	& $+0.14$	& $+0.22$	& $+0.23$	\\
		Total $(-)$			&   $-0.04$	& $-0.02$	& $-0.06$	& $-0.09$	\\
		\hline
	\end{tabular}
\end{table*}

\begin{table*}
	\caption{Variations in the \CP\ phase values $\phi_{\pm}$ (in degrees) measured for
	the $\KpiSwave\pipm$ amplitude relative to the $\KstarI\pipm$
	amplitude, and for the $\KpiSwavepm\piz$ amplitude relative to the
	$\KstarIpm\piz$ amplitude.
	}
	\label{tab:fit-cpPh-sys}
	\begin{tabular}{l||cc|cc}
		\hline
		Systematic	& \multicolumn{4}{c}{Absolute variations of
	\CP\ phase values}	\\
		\hline
				&
		\multicolumn{2}{c|}{$\KpiSwave\pipm-\KstarI\pipm$}	&
  \multicolumn{2}{c}{$\KpiSwavepm\piz-\KstarIpm\piz$} \\
  	& $\phi_{+}$	& $\phi_{-}$	& $\phi_{+}$	& $\phi_{-}$	\\
		\hline
		Self crossfeed PDFs and mapping	& $0.6$	& $4.5$	& $1.6$	& $3.1$\\
      		Correctly reconstructed and self crossfeed \NN\ PDFs    & $0.9$	& $1.6$	& $1.4$	& $2.6$ \\
		\BB\ background asymmetries & $1.4$	& $1.3$	& $0.7$	& $1.0$\\
		    Background DP PDF	& $2.5$	& $3.0$	& $2.0$	& $2.7$\\
  \BB\ background \mes, \DeltaE, \NN\ PDFs	& $0.2$	& $0.2$	& $0.2$	& $0.4$ \\
		Signal efficiency model	& $0.3$	& $1.4$	& $0.8$	& $1.2$ \\
		Fit bias		& $2.4$	& $5.2$	& $1.7$	& $0.9$\\
		$K^{*}(892)$ mass 	& $0.2$	& $0.2$	& $0.4$	& $0.6$ \\
       		$K^{*}(892)$ width	& $0.3$	& $0.3$	& $0.2$	& $0.2$ \\
		$(K\pi)^{*}_{0}$ mass  	& $6.1$	& $5.3$	& $5.9$	& $5.5$ \\
		$(K\pi)^{*}_{0}$ width 	& $4.2$	& $5.0$	& $5.2$	& $4.5$ \\
		\rhoIp\ mass 		& $0.2$	& $0.2$	& $<0.1$& $0.2$ \\
	  \rhoIp\ width 		& $0.2$	& $0.2$	& $0.1$	& $0.2$ \\
		     Blatt-Weisskopf radius 		& $1.0$	& $3.3$	& $1.3$	& $3.8$ \\
		\hline
		Total			& $8.5$	& $11.3$& $8.7$	& $9.6$ \\
		\hline
		\hline
    \multicolumn{5}{c}{Changes due to signal model} \\
		\hline
		\rhoIIp\ 			& $+3.7$ & $-0.9$ & $-3.2$ & $+9.4$ \\
		$K^{*}_{2}(1430)$		& $-2.6$ & $+0.5$ & $+3.6$ & $-8.8$	\\
		$K^{*}(1680)$			& $+1.9$ & $+3.3$ & $+2.2$ & $+14.6$\\
		\hline
		Total $(+)$			& $+4.2$ & $+0.5$ & $+4.2$ & $+17.4$\\
		Total $(-)$			& $-2.6$ & $-3.4$ & $-3.2$ & $-8.8$ \\
		\hline
	\end{tabular}
\end{table*}

\clearpage

\end{document}